\documentclass[%
 reprint,
superscriptaddress,
 amsmath,amssymb,
 aps,
prx,
]{revtex4-1}

\usepackage{graphicx}
\usepackage[dvipsnames]{xcolor}
\usepackage{amssymb}
\usepackage{amsmath}
\usepackage{amsbsy}
\usepackage{color}
\usepackage{float}

\usepackage{bm}

\usepackage{siunitx}

\begin{document}

\title{Kinetoplast DNA: nature's secret Olympic gift to polymer physicists}
\title{Kinetoplast DNA: a polymer physicist's Olympic dream}

\author{Davide Michieletto}
\affiliation{School of Physics and Astronomy, University of Edinburgh}
\affiliation{MRC Human Genetics Unit, Institute of Genetics and Cancer, University of Edinburgh}

\newcommand{\dmi}[1]{\textcolor{RoyalBlue}{#1}}

\begin{abstract}
All life forms are miraculous, but some are more inexplicable than others. Trypanosomes are by far one of the most puzzling organisms on Earth: their mitochondrial genome, also called kinetoplast DNA (kDNA) forms an Olympic-ring-like network of interlinked DNA circles, challenging conventional paradigms in both biology and physics. In this review, I will discuss kDNA from the astonished perspective of a polymer physicist and tell a story of how a single sub-celluar structure from a blood-dwelling parasite is inspiring generations of polymer chemists and physicists to create new catenated materials. 
\end{abstract}

\maketitle

\section{Introduction}
Organisms in the \textit{trypanosomatidae} family are parasitic protozoa, transmitted through the bites of tsetse flies and other blood-feeding insects, that cause significant diseases in humans and animals~\cite{Perez-Molina2018}. Trypanosomes have complex life cycles involving multiple stages and host organisms and have developed sophisticated mechanisms to evade the host's immune system. One of the most puzzling and now well studied is pan-genomic RNA editing~\cite{Simpson2000,Hajduk2010}.
The other characteristic feature of some trypanosomes is the presence of a massive DNA structure within their mitochondrion called ``kinetoplast DNA'' (kDNA), from the Greek word ``kinetikos'' (to put in motion). Organisms in the class \textit{kinetoplastida} share this distinguishing sub-cellular structure. Akin to some other mitochondrial genomes, the kDNA is made of closed, double-stranded, DNA loops, called minicircles, yet they display a truly unique feature: they are interlinked together like the rings in the Olympic flag. 

\begin{figure*}[t!]
    \centering
    \includegraphics[width=1.0\textwidth]{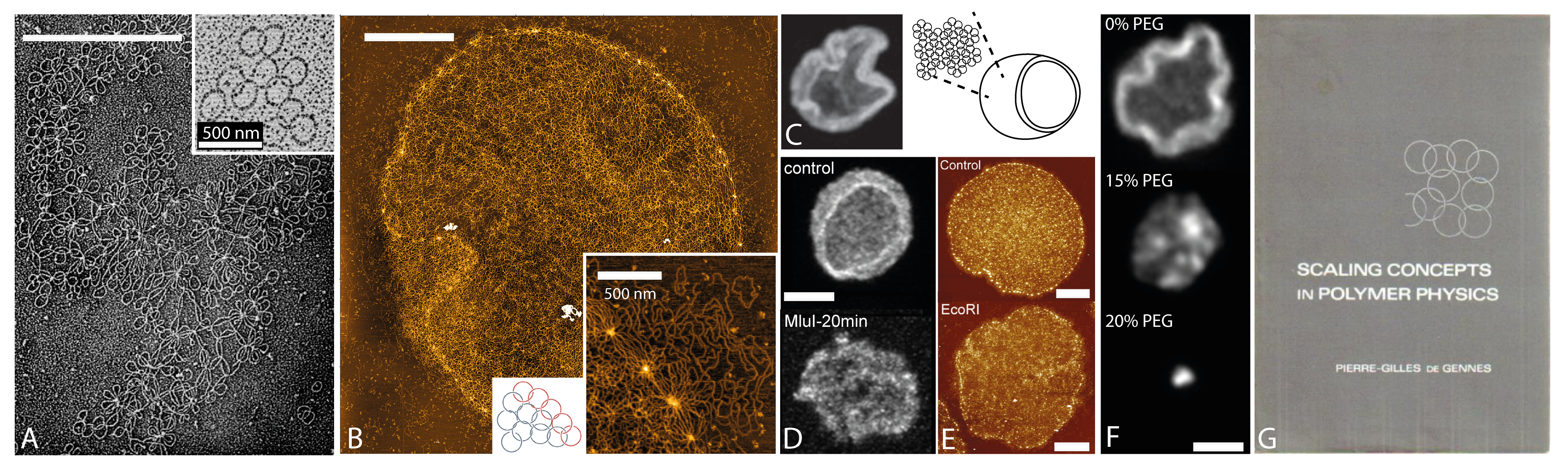}
    \vspace{-0.8 cm}
     \caption{ \textbf{A} Electron Microscopy (EM) images of fragments of kDNA from \emph{L. tarantolae}~\cite{Simpson1971}. \textbf{B} AFM image of \emph{C. fasciculata} kDNA and (large inset) zoomed in region on the rosettes at the periphery~\cite{He2023}. (Small inset) Sketch indicating the catenated structure, with red circles identifying the minicircles at the periphery. \textbf{C} Confocal image of kDNA in solution~\cite{Klotz2020}, and associated sketch. \textbf{D} Confocal images of control and partially digested kDNAs with MluI from  Ref.~\cite{Yadav2023}. \textbf{E} AFM images of control and partially digested kDNA with EcoRI from Ref.~\cite{Ramakrishnan2024}. \textbf{F} Fluorescence images of kDNA networks undergoing PEG-induced collapse~\cite{Yadav2021}. \textbf{G} Cover of P.-G. de Gennes' book ``Scaling concepts in polymer physics''~\cite{Gennes1979a} featuring catenated rings.} 
     \vspace{-0.4 cm}
    \label{fig:fig1}
\end{figure*}

Among the first kDNAs to be identified as made of interlinked DNA rings were obtained from \emph{T. cruzi} by Guy Riou~\cite{Riou1969} and from \emph{L. tarantolae} by Larry Simpson~\cite{Simpson1967,Simpson1971} as early as 1967 (Fig.~\ref{fig:fig1}A). Since then, understanding the biology and structure of the weird but beautiful kDNA became a scientific obsession for some. In fact, there isn't a ``typical'' kDNA: different Trypanosome species display different kDNA structures~\cite{Lukes2002}, ranging from a single pancake-shaped structure in \emph{C. fasciculata}, cylindrical in \emph{T. avium}, bundled in \emph{B. saltans} and dispersed in \emph{D. trypaniformis}. The shared common features are that kDNAs are formed by (i) long maxicircles encoding for genes needed for oxidative phosphorylation, (ii) short minicircles encoding for guide RNAs that perform RNA editing on the maxicircles-encoded gene transcripts and, most importantly, (iii) minicircles and maxicircles are, in most \textit{kinetoplastida} organisms, interlocked together such that the whole kDNA genome cannot be pulled apart without DNA breakage. 

The evolutionary advantage of this topologically complex arrangement has puzzled scientists for decades and some believe that it may be a case of ``ratcheted evolution'', where the same random mutations that drive higher fitness in an organism cannot be simply reversed at a later stage without increasing complexity~\cite{Borst1991}. Since kDNAs use themselves as structural templates for the synthesis of daughter networks~\cite{Perez-Morga1993,Liu2005}, major unanswered questions in the field are how did kDNAs came into existence in the first place, and how was the replication mechanism established~\cite{Shlomai1994,Liu2005,Jensen2012}. 
While there are many open questions on the biology of kDNA, there are equally many around its (bio)physics and physical properties. In this review, I will summarise the recent discoveries on the physical and topological properties of kDNAs and then I will finish with an overview of the efforts to build catenated materials directly, or indirectly, inspired by this mesmerising structure. 

\section{Physical and Topological Properties of kDNA Networks}

Different species within the class \textit{Kinetoplastida} display different kDNA structures~\cite{Lukes2002}. For example, \emph{C. fasciculata} kDNA is made of about 5000 trypanosomatidae 2.5~kb-long DNA minicircles
together with around 10 identical 30~kb-long maxicircles. It replicates by spinning on itself, like a vinyl on a turntable, meanwhile newly replicated DNA minicircles are linked to the network at special regions called ``antipodal points''~\cite{Perez-Morga1993,Perez-Morga1993b,Ferguson1994}. On the other hand, \emph{T. brucei} kDNA is made of about 10,000 genetically heterogeneous 1~kb-long minicircles and 50 identical 22~kb-long maxicircles~\cite{Stuart2005}. The genetic heterogeneity of the minicircles in both \emph{C. fasciculata} and \emph{T. brucei} depends on the extent of RNA editing required by the organism to survive in the environment. In the wild, there can be hundreds of different genetic classes of minicircles~\cite{Shapiro1995,Cooper2019}, while in lab strains these are typically reduced to tens~\cite{Ramakrishnan2024}. 

Despite species- and strain-specific variations in genetic sequences and lengths, the physical properties of kDNA networks are broadly conserved. For example, considering that the persistence length of naked DNA is around 50 nm or 150~bp -- i.e. the length at which thermal energy can significantly bend DNA -- minicircles are roughly only 6-16 persistence lengths while maxicircles are around 100-300 persistence lengths long.
The small ratio between length and persistence length renders minicircles ``semi-flexible'', meaning that the bending stiffness significantly affects the range of conformations they can assume. For instance, their semiflexible nature makes them prone to stack and to form regular, organised arrangements when placed under confinement or high concentrations~\cite{Bernabei2013,Slimani2014,Stano2022}. An organised, stacked arrangement is often seen in cryoEM sections of trypanosomes~\cite{Lukes2002} and it is thought to be facilitated by a universally shared minicircle sequence that induces a bent helical region~\cite{Silver1986}. 

The short length and semi-flexible nature of the minicircles also carries a topological consequence, i.e. that minicircles are likely to interlink only once with each of their neighbours, like the rings of a chainmail. Indeed, if they were longer or more flexible, they could wrap around each other multiple times. Topology is a property of curves that is invariant under smooth deformation, so even if the rings are deformed under thermal motion, their topology remains intact as long as the DNA strands are not broken. The number of times a minicircle is linked with another is called ``linking number'' and is mathematically formalised by the elegant Gauss double integral~\cite{Tubiana2024}.
The linking number encodes a so-called ``topological invariant'', meaning that it gives the same (integer) value irrespectively of how much two curves are deformed, as long as they are not cut, passed through each other, and glued again, for example by type II DNA Topoisomerases (topo II)~\cite{Bates2005}. Topo IIs can pass double-stranded DNA through each other by making a transient double-stranded DNA break: these enzymes essentially convert DNA molecules into so-called ``phantom'' chains that can pass through themselves. If we assume that topoIIs are not acting on them, two linked DNA minicircles will always display the same linking number. While the linking number is the number of times two minicircles wind around each-other, the ``valence'' is the number of minicircles that are linked to a certain minicircle. If we think of the kDNA as a network, the average minicircle valence represents the network connectivity. 

In 1995, Nicholas Cozzarelli and colleagues asked a simple, but very important, question: what is the topology of the kDNA network? This question cannot be answered directly via EM, as images such as the one in Fig.~\ref{fig:fig1}A do not have enough resolution to distinguish over-crossings from under-crossings and it is thus not possible to quantify the valence of the network. Only recently AFM and image analysis were able to distinguish under/over-crossings in complex DNA topologies~\cite{Gamill2024}, yet still simpler than kDNA. The intact kDNA is also too large to pass through a gel during electrophoresis, so its topology cannot be directly quantified with this method~\cite{Bates2005}. So, Cozzarelli and colleagues designed a brilliant experiment where they fragmented the kDNA network by linearising minicircles with a single cutter restriction enzyme, ran a gel, and identified the topology of small kDNA fragments made of only a handful of minicircles. From these, and assuming a fully planar and ordered lattice graph, they argued that \emph{C. fasciculata} kDNA forms a honeycomb lattice, where each minicircle is linked to another 3~\cite{Chen1995}. Recently, single-molecule Atomic Force Microscopy (AFM) coupled to molecular dynamics (MD) simulations confirmed that, on average, minicircles have valence 3, but also suggested a more random and heterogeneous topology of the network displaying, in fact, a distribution of valences~\cite{He2023} (Fig.~\ref{fig:fig1}B). This result was also independently confirmed by more recent fragmentation experiments~\cite{Ibrahim2019}. Chen, Englund and Cozzarelli went on to study different stages of replication of the kDNA in \emph{C. fasciculata} and found that in postreplicative networks, where twice as much minicircles are confined within the same mitochondrial volume, the valence builds up to (on average) 6 before returning to 3 after cell division~\cite{Chen1995a}. This implies that the role of topoIIs in kDNA is not simply to (un)link minicircles (from) to the kDNA, but it contributes to reorganise the network topology throughout replication. Being close to the percolation transition has two main evolutionary advantages: (i) most of the minicircles (nodes) in the network are linked together in a single component and (ii) individual minicircles are not redundantly linked to their neighbours. Biologically, these two properties ensure (i) reliable cell division mitigating the danger of losing genetic diversity~\cite{Borst1991} and (ii) optimal speed and efficiency of replication as individual minicircles can be unlinked from the network with the least number of decatenation events compatible with kDNA integrity~\cite{Michieletto2014a}.    
To my knowledge, a topological analysis of the kDNA network in other organisms, such as \emph{T. brucei} and \emph{L. tarantolae}, has not been carried out so we do not currently know if their minicircles have on average valence 3 in the non-replicating state.

The kDNA is physically connected to the flagellum basal body via the tripartite attachment complex (TAC), a network of proteins anchored at the mitochondrial membrane~\cite{Bonhivers2008,Amodeo2021,Kalichava2021,Hoffmann2018}. Given that the segregation of the daughter kDNAs is dependent on the integrity of the TAC and its components~\cite{Hoffmann2018}, it is broadly accepted that the interlocked DNA topology, alongside the physical attachment to the flagellum via spring-like proteins~\cite{Hoffmann2018}, contributes to faithful segregation of the daughter kDNA networks. An appealing parallel hypothesis for the existence, and persistence, of kDNA -- despite kDNA-less strains in the wild~\cite{Schnaufer2010} -- is that the kDNA network may offer an advantage to the organism movement mechanics or swimming~\footnote{To my knowledge this hypothesis was first raised and discussed in a series of exchanges between the author, Luca Tubiana and Achim Schnaufer.}. Indeed, although far from proven, the unusual interlocked topology of the kDNA may give additional physical benefits to the organisms in terms of force transduction and mechanical resilience. Intriguingly, \textit{T. brucei} strains lacking kDNA have been both found in nature and created in the lab; they are unable to develop in the insect vector as they lack the kDNA-encoded genes to perform oxidative phosphorylation~\cite{Schnaufer2010}, however no existing evidence suggests that the lack of kDNA significantly affects the movement of the parasites.    

From a material science perspective, the kDNA is a two dimensional soft material, akin to a membrane. In contrast with other naturally occurring membranes such as lipid bilayers, \emph{C. fasciculata} kDNA displays an intrinsic curvature in bulk~\cite{Klotz2020}. Specifically, it assumes a cup (or ``shower cap'') shape with positive Gaussian curvature~\cite{Klotz2020,Soh2021,Polson2021} (see Fig.~\ref{fig:fig1}C) and an overall diameter of about 5 $\mu$m (its diameter within the mitochondrion being $\simeq$ 1 $\mu$m). It is generally accepted that the intrinsic curvature is due to its ``edge-loop'': a line of rosette-like structures surrounding the network that displays redundant links~\cite{He2023,Ragotskie2024} (Fig.~\ref{fig:fig1}B, inset). Although the effect of the minicircle linking chirality on the macroscopic intrinsic curvature of the catenated membrane is another fascinating hypothesis (see Refs.~\cite{Polson2021,Tubiana2022,Klotz2024chirality}). 
The edge-loop in \emph{C. fasciculata} kDNA is likely formed during its replication, which occurs under confinement and constant area (and fixed perimeter)~\cite{Perez-Morga1993,Chen1995a}. When the kDNA is extracted from the mitochondrion and placed in bulk, its area expands to maximise entropy and minimise steric interactions between the minicircles, but the maximum extent of the perimeter is bounded by the finite number of minicircles and their contour length. For example, in the inset of Fig.~\ref{fig:fig1}B one can appreciate the presence of ``rosettes'' connected by strands of DNA minicircles that are pulled taut due to the full absoprtion of the kDNA on the mica surface.  Once lifted from the surface, these stretched DNA strands act as springs, shrinking the kDNA perimeter. These two effects, expansion of the area and finite extension of the perimeter, induce buckling and shapes with positive Gaussian curvature~\cite{He2023,Klotz2020}.  

Membranes also possess intrinsic bending stiffness $\kappa$. Klotz directly measured \emph{C. fasciculata} kDNA bending stiffness by using a microfluidic device that induced an elongation and sudden release of individual networks~\cite{Klotz2020}. In analogy with spherical vesicles, the bending stiffness could be then estimated through the relaxation time of the deformation $\tau$, as $\kappa = \eta r^3/ (\pi \tau)$, where $r$ is the size of the kDNA at equilibrium and $\eta$ the viscosity of the solvent. This calculation yields $\kappa \simeq 1.9$ $10^{-19}$ J, which is very similar to that of standard lipid membranes~\cite{Klotz2020}. On the other hand, membranes display in-plane and out-of-plane deformation ``modes'' corresponding to pulling by the edges, or buckling, a piece of paper. These deformations follow, \textit{a priori}, different stiffness~\cite{Shankar2021}, e.g. when you push a piece of paper on the edges it will prefer to buckle rather than reduce its area. Indirect estimations based on AFM images suggest that kDNAs may display an ``ultra-low'' in-plane stiffness due to the very low packing DNA density when outside the mithocondrion and to the subisostatic nature of the network~\cite{He2023}: once expanded, the minicircles swell while remaining connected via topological links. In-plane compression of such structure mostly pushes against entropy of the minicircles rather than steric interactions and thus allow for a large deformation before yielding and buckling out-of-plane. Individual minicircles may be thought of as springs, resisting both stretching and compression and they combine together to give an in-plane Young modulus $Y \simeq 0.1$ pN/$\mu$m~\cite{He2023}. The bending (out-of-plane) stiffness can then be estimated as $\kappa \simeq 3 \times 10^{-21}$ J. Both of these are 2-3 orders of magnitude smaller than traditional lipid bilayers, meaning that the kDNA would be extremely easy to deform, stretch and bend. It is likely that these numbers will differ in kDNAs from other organisms, as they depend on the size of the minicircles and the DNA density.
Additionally, it is natural to expect that kDNA's material properties will be affected by topology of the network itself. Recent \emph{in vitro} imaging~\cite{Yadav2023} (see Fig.~\ref{fig:fig1}D) and AFM~\cite{Ramakrishnan2024} experiments (Fig.~\ref{fig:fig1}E) done on partially digested \emph{C. fasciculata} kDNAs reported significant changes when subsets of the maxi or minicircles were selectively removed from the network. Specifically, cleavage of maxicircles yielded significant shrinking of the networks when adsorbed on the mica, while linearisation of both maxicircles and the first minor minicircle class (around 15\% of the total kDNA mass) yielded significantly deformed, though wider, networks~\cite{Ramakrishnan2024}  (Fig.~\ref{fig:fig1}E). These observations can be understood via a polymer physics argument, as the area of an absorbed network follows $(R/R_0)^2 \sim N/\kappa$, i.e. directly proportional to the amount of mass in the network ($N$), but inversely proportional to its stiffness ($\kappa$). In turn, the smaller area of maxicircles-less kDNAs suggest that the network loses mass but $\kappa$ remains mostly constant, while the wider area of EcoRI-treated kDNAs suggests that removing minicircles significantly reduces the stiffness $\kappa$~\cite{Ramakrishnan2024}. These observations are in line with confocal imaging done on kDNAs in bulk~\cite{Yadav2023}, where the authors measured the autocorrelation of the shape of the networks and found that increasing DNA digestion leads to network softening (Fig.~\ref{fig:fig1}D). 

Not only are kDNAs the archetypes of catenated networks, but they also represent rare, natural examples of 2D polymers. For this reason, \emph{C. fasciculata} kDNA was also recently used to study confinement and ionic and polymer-salt-induced ($\psi$) transition of two-dimensional polymers~\cite{Soh2020,Soh2021,Yadav2021}. It was found that $\psi$ transition in kDNAs is smoother than the first-order transition seen in linear (``one dimensional'') polymers and it displays a coexistence of kDNAs with different shapes and sizes (Fig.~\ref{fig:fig1}F). Since these conformations appear to be long-lived, Soh and coauthors argued that they may reflect an underlying ``rugged'' energy landscape, in contrast with the simpler double-well free energy landscape observed for linear polymers~\cite{Yadav2021}. Given the challenge in synthesising truly two dimensional polymers, kDNAs thus represents a unique opportunity for polymer physicists to study the properties and scaling laws of such polymers.

\begin{figure*}[t!]
    \centering
    \includegraphics[width=1.0\textwidth]{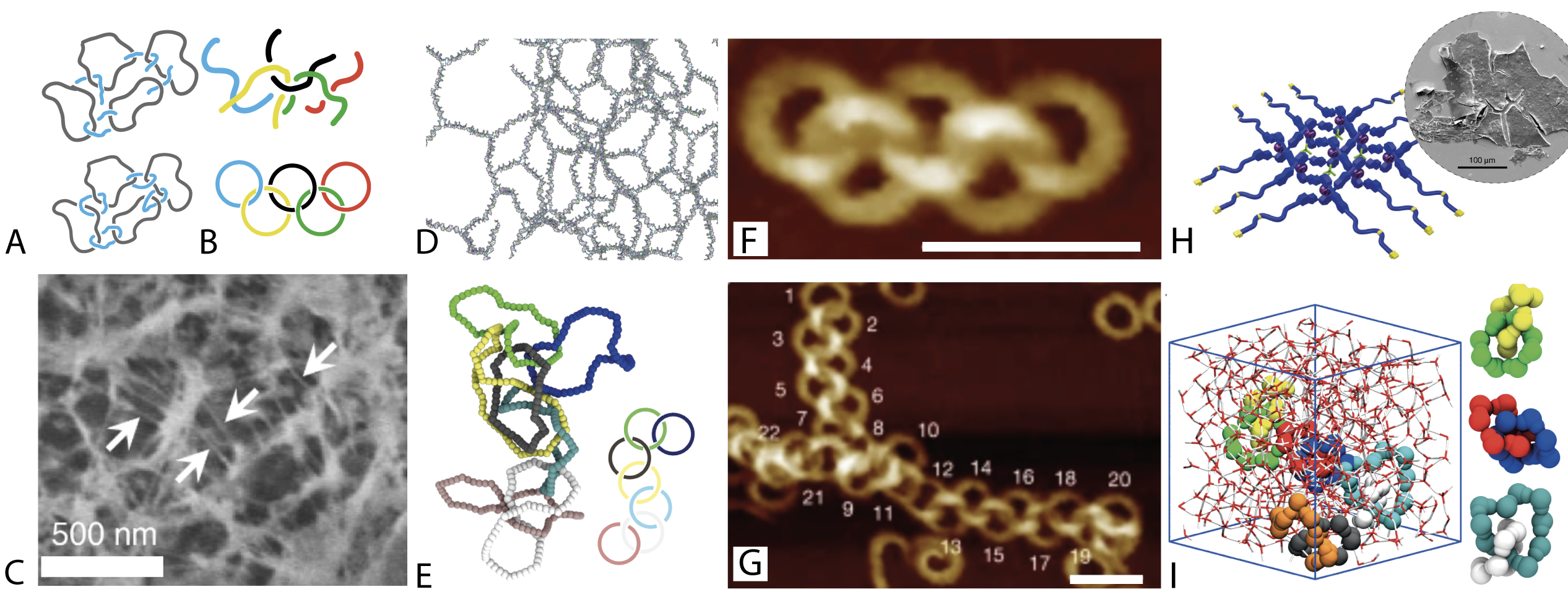}
    \vspace{-0.8 cm}
     \caption{ \textbf{A-B} Different pathways to assemble an Olympic network: \textbf{A} de Gennes construction~\cite{Raphael1997}, \textbf{B} Pickett's ``DNA origami'' construction~\cite{Pickett2006}. \textbf{C} Cryo-SEM image of an Olympic gel assembled via a DNA library~\cite{Speed2024}. \textbf{D} OxDNA simulation of a hydrogel structure formed by DNA nanostars~\cite{Palombo2023}. \textbf{E} Polycatenated looped motifs in DNA nanostar hydrogels~\cite{Palombo2023}. \textbf{F-G} Polycatenated synthetic structures from Ref.~\cite{Datta2020} (scale bars 50 nm). \textbf{H} Two-dimensional woven structures~\cite{August2020}. \textbf{I} Catenated motifs in simulated high-density amorphous ice~\cite{Fosado2024ice}.} 
     \vspace{-0.4 cm}
    \label{fig:fig2}
\end{figure*}

\section{An Olympic Dream}

I vividly remember my amazement when I realised that the micrograph picture of \emph{Leshmania tarantolae} (\emph{L. tarantolae}) kDNA from Larry Simpson~\cite{Simpson1971} (Fig.~\ref{fig:fig1}A) was essentially identical to the cover of the 1979 book ``Scaling concepts in Polymer Physics'' by the Nobel Prize Laureate in Physics Pierre-Gilles de Gennes~\cite{Gennes1979a} (Fig.~\ref{fig:fig1}G). 
In fact, there is no evidence that de Gennes knew about kDNAs. Using his imagination, he came up with a completely new class of polymer gels where topological crosslinks would replace traditional chemical ones. He dubbed such an imaginary structure an ``Olympic gel'', and described a strategy to realise it in the lab~\cite{Gennes1979a,Raphael1997}. The so-called ``de Gennes's construction'' involves mixing a high concentration of large unlinked ring polymers with a smaller concentration of linear chains with reactive ends. During the gelation reaction, the linear chains would loop onto themselves and linking with the already present large rings~\cite{Raphael1997} (Fig.~\ref{fig:fig2}A). It is interesting that de Gennes never suggested topo II-like strand passage reactions (Fig.~\ref{fig:fig2}B), which indicates he was unaware of this protein or that he did not realise DNA could be used to make Olympic networks. To my knowledge, the de Gennes' construction has never been actually implemented in the lab. Indeed, one of the major challenges for this strategy to work is the competition between cyclisation and polymer condensation~\cite{Lang2021,Panoukidou2024}. To form a system-spanning network of interlinked loops, high concentrations near or above chain overlapping are needed to favour concatenation. At these high concentrations (and due to the low looping probability of long chains~\cite{Shimada1984}), unspecific reactions between polymer ends favour condensation (or polymerisation), i.e. the formation of long linear polymeric strands, instead of cyclisation~\cite{Panoukidou2024}. 

The polymerisation can be reduced, or avoided altogether, by the design of linear chains with specific end groups such that closure occurs only between ends belonging to the same ``family''~\cite{Pickett2006,Fischer2015}. Dividing the stock solution of reactive chains in $B$ orthogonal families (or ``colors'') would reduce the probability of two polymers from the same family being within the same volume of each other by a factor $B$. The challenge with this approach is that one would need a large number $B$ of orthogonal and specific end groups to successfully assemble an Olympic gel~\cite{Pickettopo II006} (Fig.~\ref{fig:fig2}C). Reaching such high number of orthogonal functional groups is unfeasible with traditional synthetic chemistry. On the other hand, Watson-Crick base pairing of DNA sequences with enough variation would endow the groups with the needed specificity. An Olympic gel assembled by a combinatorial DNA library of more than 16,000 clones was recently reported using this method~\cite{Speed2024} (see Fig.~\ref{fig:fig2}D) for a cryo-SEM image of the inner structure of such gel). The library was constructed by cloning a ``lock-and-key'' domain made of 16 random oligonucleotides -- yielding $4^{16}$ different sequences -- and flanked by nicking sites into vectors which were then transformed into \emph{E. coli}. After amplification and nicking, the clones could be opened by raising the temperature above the melting temperature of the ``lock-and-key'' domain, thereby allowing interpenetration, and subsequently closed specifically by cooling down the solution. Even at very high concentrations, the chances for any one of these reactive ends to find another with the same sequence is negligible, and therefore the formation of closed DNA circle is the most probable outcome, in turn yielding a concatenated gel above a certain concentration~\cite{Speed2024}. This is an example of how cloning techniques from molecular biology can help the design of materials that physicists could only dream of a few decades ago.    

Thanks to these interdisciplinary interactions, physicists also recently realised that another way to form an Olympic gel is by using topo IIs. At high enough plasmid concentration, topo II can form links and knots in solutions of DNA~\cite{Kreuzer1980}. This mechanism was used by Kim~\cite{Kim2013} and Krajina~\cite{Krajina2018} to study concatenated fluids of DNA plasmids: they observed that these solutions were fluid-like when topo II and ATP were present in the system but gelled when ICRF-193 (a topo II inhibitor) or EDTA were added. More specifically, dense solutions of moderately long and nicked plasmid behaved as viscoelastic liquids (like silly putty) in presence of topo II: fluid on timescales either, longer than the topo II-catalysed strand crossing reaction, or shorter than the time for two catenated plasmids to feel the effect of topological constraints, but solid-like in between these two timescales~\cite{Krajina2018}. After adding EDTA~\cite{Krajina2018} or ICRF-193~\cite{Kim2013} the fluids lost the long-time fluid-like behaviour and instead acquired a solid-like behaviour with an elastic plateau that characterises permanently crosslinked networks. The downside of this assembly pathway is that topo II itself may work as a crosslink between two DNA strands, especially in the presence of drugs~\cite{Le2022}, which may affect the mechanical properties of the network. Computer simulations suggest that dynamic topo II-like strand-crossing reactions can yield polymer solutions that relax slower than their uncatenated counterparts by suitably tuning the rate of strand crossing reactions~\cite{Ubertini2021}. This could be realised experimentally by tuning the concentration of topo II in solution so that the timescale of strand crossing is longer than the self-diffusion time of the DNA plasmids. To my knowledge, this strategy of tuning the viscoelasticty of DNA solutions has not been experimentally tested yet.

Olympic gels are interesting for polymer physicists and material scientists because, once their topology is quenched (or fixed), they are expected to display exotic material and swelling properties. For instance, at small strains, the force $f$ needed to deform an Olympic gel is predicted to be insensitive on the strain, thanks to the fact that topological links allow the rings to move freely for small deformations $\lambda$; however, at large strains a non-linear behaviour is expected where $f \sim \lambda^{2/3}$~\cite{Vilgis1997}. In recent simulations, Lang measured a stress scaling as $f \sim \lambda^{0.78}$, close to the theoretical prediction~\cite{Speed2024}. Olympic gels are also expected to display unusual swelling: if during preparation some overlapping rings are not catenated, they are then freely redistributed, causing large (non-affine) swelling, also called ``disinterpersion''~\cite{Lang2014a}. Larger preparation concentrations of polymers ($\phi_0)$ or larger polymerisation index ($N$) have been empirically shown in computer simulations~\cite{Lang2014a} to reduce the swelling ratio $Q$ as $Q \sim N^{-0.28} \phi_0^{-0.72}$; this is in contrast with the classic behaviour for chemically crosslinked gels scaling as $Q \sim N^{0.57} \phi_0^{-1/4}$~\cite{Rubinsteinbook,Lang2014a} (note the change in sign of the exponent associated with polymerisation index $N$). The recent Olympic gel prepared in Ref.~\cite{Speed2024} is in qualitative agreement with these theories, but more work is needed to fully test and understand these predictions. 
Finally, the elastic modulus of the Olympic gel in Ref.~\cite{Speed2024} has been found to scale as a function of the DNA concentration as $G_p \sim \phi_0^{\alpha}$, with $\alpha$ close to the exponent of $2.3$ predicted for entangled polymers~\cite{Rubinsteinbook}, yet also close to $2.5$ found in DNA nanostar hydrogels, where the elasticity is determined by the degree of catenation between looped motifs in the network structure~\cite{Palombo2023} (Fig.~\ref{fig:fig2}D-E). Now that it is possible to prepare Olympic gels in the lab through DNA plasmid libraries, I believe that it will be very exciting for the polymer physics community to precisely understand the origin of its elastic behaviour.

\section{Other Catenated Materials}

Physicists are not the only ones dreaming of catenated structures. Around the same time that de Gennes dreamed of the Olympic gels, chemist Jean-Pierre Sauvage and colleagues published~\footnote{There is no evidence that de Gennes and Sauvage knew of each other's work, despite being compatriots.} the first synthetic molecular catenane in 1983~\cite{Sauvage1983} citing prior work from Vinograd -- who discovered in 1967 linked DNA plasmids in the mitochondrion of HeLa cells~\cite{Hudson1967,Clayton1967} -- and of Cozzarelli -- who had recently discovered (in 1980) the formation and dissolution of DNA catenane by topo IIs~\cite{Kreuzer1980}. Differently from physicists, chemists did know of the work of molecular biologists on catenated DNA, but they were likely unaware of kDNA too.  

Since the work of Sauvage, and of fellow chemists Sir Fraser Stoddard and Bernard Feringa, molecular catenanes, mechanically interlocked molecules and, more generally, molecular topology, has become a field of chemistry in itself~\cite{Fielden2017,Schaufelberger2020}. Complex knotting and linking at the microscale have been developed in chemistry labs, mostly unaware of the existence of kDNA. For chemists, complex topologies enhance the functionality of synthetic supramolecuar structures, but they are challenging to synthetise in a rational and reproducible manner: synthetic chemistry does not have the equivalent of a topo II, which is what trypanosomes heavily rely on to assemble and replicate kDNA. Despite this, chemists have made a range of impressive structures, ranging from nano-polyrotaxanes~\cite{Datta2020} (Fig.~\ref{fig:fig1}F-G), mechanically interlocked networks with rotaxane crosslinks~\cite{Zhao2021}, 2D weavings~\cite{August2020} (Fig.~\ref{fig:fig2}H) and metamaterials~\cite{Kadic2019}. Most of these mesmerising structures rely on metal-coordination chemistry to drive precise assembly~\cite{Fielden2017} and have thus achieved degrees of control not present in the catenated and knotted DNA structures~\cite{Hudson1967,Kreuzer1980} -- the ones that originally inspired Sauvage and coworkers -- which are instead mostly driven by entropy. Entropy should also play a major role also in the assembly of kDNA, as it is unlikely that organisms rely on precise coordination of each individual minicircle to assemble the network. 

The realisation that the kDNA was biology's answer on how to make Olympic networks started to percolate in the mathematics, physics and chemistry communities only in the last 10 years~\cite{Diao2012,Michieletto2014a,Schmidt2011}. This cross-fertilisation rekindled a broader interest for catenated materials. Since then, a considerable amount of work has been done to study polycatenanes~\cite{Rauscher2020,Tubiana2022}, interlinked 2D networks~\cite{Luengo-Marquez2024,Polson2021,Klotz2024,Polles2016} and 3D Olympic gels~\cite{Speed2024,Lang2014a,Ubertini2021,Krajina2018}. In parallel, researchers also managed to create Olympic-like structures using strategies from the quickly growing field of DNA origami, such as 2D auxetic structures~\cite{Li2021}, meta-DNA~\cite{Yao2020},  interlocked~\cite{List2016,Valero2019}, and catenaned~\cite{Schmidt2011,Peil2020} DNA origami. So-called DNA ``nanostars'' have also been used to realised viscoelastic fluids and hydrogels~\cite{Biffi2013,Conrad2019a} and these soft materials were recently found to display an elastic behaviour that is dictated by the catenation of looped motifs within the network structure~\cite{Palombo2023} (Fig.~\ref{fig:fig2}D-E).  More generally, the role of topology in materials, complex fluids and biology, e.g. DNA and proteins, is gaining considerable attention from the physics community~\cite{Tubiana2024}. This field has recently received a further kick by the discovery that topologically complex structures characterise thermodynamic phase transitions in amorphous liquids, such as water~\cite{Neophytou2022} and ice~\cite{Fosado2024ice} (Fig.~\ref{fig:fig2}I). Given how generic these amorphous structures are, similar effects are expected in many other fluids. For these reasons -- and thanks to the universal power of topology which can describe extremely diverse objects, from molecular knots to solar flares~\cite{Tubiana2024} -- studying the self-assembly pathway and material properties of the archetypal kDNA in trypanosomes may lead to a better understanding of completely different forms of liquids and materials.

Finally, far from a theoretical quirk, kDNAs and catenated structures are finding extensive applications in drug discovery. Indeed, kDNA substrates can be used to assay topo IIs~\cite{Nitiss1998} as their strand-crossing ability is necessary to release minicircles from the kDNA. Decatenation assays have been tremendously useful in mechanistic studies of topo IIs and also in the discovery of antibiotics and anti-cancer drugs targeted to type II topos.  Recently, a singly-linked substrate for such reactions (bis-cat) has been developed~\cite{Waraich2020}, which may prove to be more convenient than kDNA for these purposes.

\section{Conclusions}
Olympic structures made of interlocked rings are highly sought-after in physics, chemistry and material science thanks to their expected unusual and exotic material properties, yet they proved very challenging to realise. It is humbling and at the same time exciting to realise that de Gennes' dream of creating materials made of interlinked loops had been ``invented'' by nature millions of years earlier. It is thus not surprising that the discovery of kDNA networks in the 1960s, and the recent wave of interest from the physics community, is rekindling the passion for topologically complex materials, and especially for those employing DNA as a building block, such as DNA origami and hydrogels. Catenated DNA structures have the advantage that they maintain a large local DNA density, whilst the individual rings have a large degree of freedom and can thus be easily accessed by proteins and enzymes. In turn, it is expected that kDNA-like structures may be an ideal substrate for rolling circle amplification for large scale production of DNA, or cell-free protein synthesis. Thus, studying how are kDNAs assembled and how are their topology regulated \emph{in vivo} may not only lead to profound insights into their biogenesis and function but also into how we can make new tunable, and scalable, biotechnologies based on catenated DNA structures in the future. 

\section{Acknowledgements}
DM acknowledges the Royal Society and the European Research Council (grant agreement No 947918, TAP) for funding. The authors also acknowledge the contribution of the COST Action Eutopia, CA17139. DM thanks Tony Maxwell, Achim Schnaufer, Larry Simpson, Joaquim Roca, Luca Tubiana and Antonio Valdes for stimulating conversations and thoughtful suggestions on the manuscript. For the purpose of open access, the author has applied a Creative Commons Attribution (CC BY) licence to any Author Accepted Manuscript version arising from this submission.

\bibliography{biblio}

\begin{thebibliography}{90}%
\makeatletter
\providecommand \@ifxundefined [1]{%
 \@ifx{#1\undefined}
}%
\providecommand \@ifnum [1]{%
 \ifnum #1\expandafter \@firstoftwo
 \else \expandafter \@secondoftwo
 \fi
}%
\providecommand \@ifx [1]{%
 \ifx #1\expandafter \@firstoftwo
 \else \expandafter \@secondoftwo
 \fi
}%
\providecommand \natexlab [1]{#1}%
\providecommand \enquote  [1]{``#1''}%
\providecommand \bibnamefont  [1]{#1}%
\providecommand \bibfnamefont [1]{#1}%
\providecommand \citenamefont [1]{#1}%
\providecommand \href@noop [0]{\@secondoftwo}%
\providecommand \href [0]{\begingroup \@sanitize@url \@href}%
\providecommand \@href[1]{\@@startlink{#1}\@@href}%
\providecommand \@@href[1]{\endgroup#1\@@endlink}%
\providecommand \@sanitize@url [0]{\catcode `\\12\catcode `\$12\catcode
  `\&12\catcode `\#12\catcode `\^12\catcode `\_12\catcode `\%12\relax}%
\providecommand \@@startlink[1]{}%
\providecommand \@@endlink[0]{}%
\providecommand \url  [0]{\begingroup\@sanitize@url \@url }%
\providecommand \@url [1]{\endgroup\@href {#1}{\urlprefix }}%
\providecommand \urlprefix  [0]{URL }%
\providecommand \Eprint [0]{\href }%
\providecommand \doibase [0]{http://dx.doi.org/}%
\providecommand \selectlanguage [0]{\@gobble}%
\providecommand \bibinfo  [0]{\@secondoftwo}%
\providecommand \bibfield  [0]{\@secondoftwo}%
\providecommand \translation [1]{[#1]}%
\providecommand \BibitemOpen [0]{}%
\providecommand \bibitemStop [0]{}%
\providecommand \bibitemNoStop [0]{.\EOS\space}%
\providecommand \EOS [0]{\spacefactor3000\relax}%
\providecommand \BibitemShut  [1]{\csname bibitem#1\endcsname}%
\let\auto@bib@innerbib\@empty
\bibitem [{\citenamefont {P{\'{e}}rez-Molina}\ and\ \citenamefont
  {Molina}(2018)}]{Perez-Molina2018}%
  \BibitemOpen
  \bibfield  {author} {\bibinfo {author} {\bibfnamefont {J.~A.}\ \bibnamefont
  {P{\'{e}}rez-Molina}}\ and\ \bibinfo {author} {\bibfnamefont
  {I.}~\bibnamefont {Molina}},\ }\href {\doibase 10.1016/S0140-6736(17)31612-4}
  {\bibfield  {journal} {\bibinfo  {journal} {The Lancet}\ }\textbf {\bibinfo
  {volume} {391}},\ \bibinfo {pages} {82} (\bibinfo {year} {2018})}\BibitemShut
  {NoStop}%
\bibitem [{\citenamefont {Simpson}\ \emph {et~al.}(2000)\citenamefont
  {Simpson}, \citenamefont {Thiemann}, \citenamefont {Savill}, \citenamefont
  {Alfonzo},\ and\ \citenamefont {Maslov}}]{Simpson2000}%
  \BibitemOpen
  \bibfield  {author} {\bibinfo {author} {\bibfnamefont {L.}~\bibnamefont
  {Simpson}}, \bibinfo {author} {\bibfnamefont {O.~H.}\ \bibnamefont
  {Thiemann}}, \bibinfo {author} {\bibfnamefont {N.~J.}\ \bibnamefont
  {Savill}}, \bibinfo {author} {\bibfnamefont {J.~D.}\ \bibnamefont {Alfonzo}},
  \ and\ \bibinfo {author} {\bibfnamefont {D.~A.}\ \bibnamefont {Maslov}},\
  }\href@noop {} {\bibfield  {journal} {\bibinfo  {journal} {Proc. Natl. Acad.
  Sci. USA}\ }\textbf {\bibinfo {volume} {97}},\ \bibinfo {pages} {6986}
  (\bibinfo {year} {2000})}\BibitemShut {NoStop}%
\bibitem [{\citenamefont {Hajduk}\ and\ \citenamefont
  {Ochsenreiter}(2010)}]{Hajduk2010}%
  \BibitemOpen
  \bibfield  {author} {\bibinfo {author} {\bibfnamefont {S.}~\bibnamefont
  {Hajduk}}\ and\ \bibinfo {author} {\bibfnamefont {T.}~\bibnamefont
  {Ochsenreiter}},\ }\href {\doibase 10.4161/rna.7.2.11393} {\bibfield
  {journal} {\bibinfo  {journal} {RNA Biology}\ }\textbf {\bibinfo {volume}
  {7}},\ \bibinfo {pages} {229} (\bibinfo {year} {2010})}\BibitemShut {NoStop}%
\bibitem [{\citenamefont {Riou}\ and\ \citenamefont {Delain}(1969)}]{Riou1969}%
  \BibitemOpen
  \bibfield  {author} {\bibinfo {author} {\bibfnamefont {G.}~\bibnamefont
  {Riou}}\ and\ \bibinfo {author} {\bibfnamefont {E.}~\bibnamefont {Delain}},\
  }\href {\doibase 10.1073/pnas.62.1.210} {\bibfield  {journal} {\bibinfo
  {journal} {Proceedings of the National Academy of Sciences of the United
  States of America}\ }\textbf {\bibinfo {volume} {62}},\ \bibinfo {pages}
  {210} (\bibinfo {year} {1969})}\BibitemShut {NoStop}%
\bibitem [{\citenamefont {Simpson}(1967)}]{Simpson1967}%
  \BibitemOpen
  \bibfield  {author} {\bibinfo {author} {\bibfnamefont {L.~P.}\ \bibnamefont
  {Simpson}},\ }\href@noop {} {\bibfield  {journal} {\bibinfo  {journal} {Atlas
  de Symposia sobre a Biota Amazonica (Pathologia)}\ }\textbf {\bibinfo
  {volume} {6}},\ \bibinfo {pages} {231} (\bibinfo {year} {1967})}\BibitemShut
  {NoStop}%
\bibitem [{\citenamefont {Simpson}\ and\ \citenamefont {{Da
  Silva}}(1971)}]{Simpson1971}%
  \BibitemOpen
  \bibfield  {author} {\bibinfo {author} {\bibfnamefont {L.}~\bibnamefont
  {Simpson}}\ and\ \bibinfo {author} {\bibfnamefont {A.}~\bibnamefont {{Da
  Silva}}},\ }\href
  {http://www.sciencedirect.com/science/article/pii/0022283671903949}
  {\bibfield  {journal} {\bibinfo  {journal} {J. Mol. Biol.}\ }\textbf
  {\bibinfo {volume} {56}},\ \bibinfo {pages} {443} (\bibinfo {year}
  {1971})}\BibitemShut {NoStop}%
\bibitem [{\citenamefont {Lukes}\ \emph {et~al.}(2002)\citenamefont {Lukes},
  \citenamefont {Guilbride},\ and\ \citenamefont {Vot{\'{y}}pka}}]{Lukes2002}%
  \BibitemOpen
  \bibfield  {author} {\bibinfo {author} {\bibfnamefont {J.}~\bibnamefont
  {Lukes}}, \bibinfo {author} {\bibfnamefont {D.}~\bibnamefont {Guilbride}}, \
  and\ \bibinfo {author} {\bibfnamefont {J.}~\bibnamefont {Vot{\'{y}}pka}},\
  }\href {\doibase 10.1128/EC.1.4.495} {\bibfield  {journal} {\bibinfo
  {journal} {Eukaryot. Cell}\ }\textbf {\bibinfo {volume} {1}},\ \bibinfo
  {pages} {495} (\bibinfo {year} {2002})}\BibitemShut {NoStop}%
\bibitem [{\citenamefont {Borst}(1991)}]{Borst1991}%
  \BibitemOpen
  \bibfield  {author} {\bibinfo {author} {\bibfnamefont {P.}~\bibnamefont
  {Borst}},\ }\href
  {http://www.sciencedirect.com/science/article/pii/016895259190374Y}
  {\bibfield  {journal} {\bibinfo  {journal} {Trends Genet.}\ }\textbf
  {\bibinfo {volume} {7}} (\bibinfo {year} {1991})}\BibitemShut {NoStop}%
\bibitem [{\citenamefont {P{\'{e}}rez-Morga}\ and\ \citenamefont
  {Englund}(1993)}]{Perez-Morga1993}%
  \BibitemOpen
  \bibfield  {author} {\bibinfo {author} {\bibfnamefont {D.~L.}\ \bibnamefont
  {P{\'{e}}rez-Morga}}\ and\ \bibinfo {author} {\bibfnamefont {P.~T.}\
  \bibnamefont {Englund}},\ }\href {http://www.ncbi.nlm.nih.gov/pubmed/8395351}
  {\bibfield  {journal} {\bibinfo  {journal} {Cell}\ }\textbf {\bibinfo
  {volume} {74}},\ \bibinfo {pages} {703} (\bibinfo {year} {1993})}\BibitemShut
  {NoStop}%
\bibitem [{\citenamefont {Liu}\ \emph {et~al.}(2005)\citenamefont {Liu},
  \citenamefont {Liu}, \citenamefont {Motyka}, \citenamefont {Agbo},\ and\
  \citenamefont {Englund}}]{Liu2005}%
  \BibitemOpen
  \bibfield  {author} {\bibinfo {author} {\bibfnamefont {B.}~\bibnamefont
  {Liu}}, \bibinfo {author} {\bibfnamefont {Y.}~\bibnamefont {Liu}}, \bibinfo
  {author} {\bibfnamefont {S.~a.}\ \bibnamefont {Motyka}}, \bibinfo {author}
  {\bibfnamefont {E.~E.~C.}\ \bibnamefont {Agbo}}, \ and\ \bibinfo {author}
  {\bibfnamefont {P.~T.}\ \bibnamefont {Englund}},\ }\href {\doibase
  10.1016/j.pt.2005.06.008} {\bibfield  {journal} {\bibinfo  {journal} {Trends
  in parasitology}\ }\textbf {\bibinfo {volume} {21}},\ \bibinfo {pages} {363}
  (\bibinfo {year} {2005})}\BibitemShut {NoStop}%
\bibitem [{\citenamefont {Shlomai}(1994)}]{Shlomai1994}%
  \BibitemOpen
  \bibfield  {author} {\bibinfo {author} {\bibfnamefont {J.}~\bibnamefont
  {Shlomai}},\ }\href {http://www.ncbi.nlm.nih.gov/pubmed/15275411} {\bibfield
  {journal} {\bibinfo  {journal} {Parasitology today (Personal ed.)}\ }\textbf
  {\bibinfo {volume} {10}},\ \bibinfo {pages} {341} (\bibinfo {year}
  {1994})}\BibitemShut {NoStop}%
\bibitem [{\citenamefont {Jensen}\ and\ \citenamefont
  {Englund}(2012)}]{Jensen2012}%
  \BibitemOpen
  \bibfield  {author} {\bibinfo {author} {\bibfnamefont {R.~E.}\ \bibnamefont
  {Jensen}}\ and\ \bibinfo {author} {\bibfnamefont {P.~T.}\ \bibnamefont
  {Englund}},\ }\href {\doibase 10.1146/annurev-micro-092611-150057} {\bibfield
   {journal} {\bibinfo  {journal} {Annu. Rev. Microbiol.}\ }\textbf {\bibinfo
  {volume} {66}},\ \bibinfo {pages} {473} (\bibinfo {year} {2012})}\BibitemShut
  {NoStop}%
\bibitem [{\citenamefont {He}\ \emph {et~al.}(2023)\citenamefont {He},
  \citenamefont {Katan}, \citenamefont {Tubiana}, \citenamefont {Dekker},\ and\
  \citenamefont {Michieletto}}]{He2023}%
  \BibitemOpen
  \bibfield  {author} {\bibinfo {author} {\bibfnamefont {P.}~\bibnamefont
  {He}}, \bibinfo {author} {\bibfnamefont {A.~J.}\ \bibnamefont {Katan}},
  \bibinfo {author} {\bibfnamefont {L.}~\bibnamefont {Tubiana}}, \bibinfo
  {author} {\bibfnamefont {C.}~\bibnamefont {Dekker}}, \ and\ \bibinfo {author}
  {\bibfnamefont {D.}~\bibnamefont {Michieletto}},\ }\href
  {http://arxiv.org/abs/2209.01293} {\bibfield  {journal} {\bibinfo  {journal}
  {Physical Review X}\ }\textbf {\bibinfo {volume} {13}},\ \bibinfo {pages} {1}
  (\bibinfo {year} {2023})}\BibitemShut {NoStop}%
\bibitem [{\citenamefont {Klotz}\ \emph {et~al.}(2020)\citenamefont {Klotz},
  \citenamefont {Soh},\ and\ \citenamefont {Doyle}}]{Klotz2020}%
  \BibitemOpen
  \bibfield  {author} {\bibinfo {author} {\bibfnamefont {A.~R.}\ \bibnamefont
  {Klotz}}, \bibinfo {author} {\bibfnamefont {B.~W.}\ \bibnamefont {Soh}}, \
  and\ \bibinfo {author} {\bibfnamefont {P.~S.}\ \bibnamefont {Doyle}},\
  }\href@noop {} {\bibfield  {journal} {\bibinfo  {journal} {Proceedings of the
  National Academy of Sciences of the United States of America}\ }\textbf
  {\bibinfo {volume} {117}},\ \bibinfo {pages} {121} (\bibinfo {year}
  {2020})}\BibitemShut {NoStop}%
\bibitem [{\citenamefont {Yadav}\ \emph {et~al.}(2023)\citenamefont {Yadav},
  \citenamefont {{Al Sulaiman}},\ and\ \citenamefont {Doyle}}]{Yadav2023}%
  \BibitemOpen
  \bibfield  {author} {\bibinfo {author} {\bibfnamefont {I.}~\bibnamefont
  {Yadav}}, \bibinfo {author} {\bibfnamefont {D.}~\bibnamefont {{Al
  Sulaiman}}}, \ and\ \bibinfo {author} {\bibfnamefont {P.~S.}\ \bibnamefont
  {Doyle}},\ }\href@noop {} {\bibfield  {journal} {\bibinfo  {journal}
  {Physical Review Research}\ }\textbf {\bibinfo {volume} {5}},\ \bibinfo
  {pages} {1} (\bibinfo {year} {2023})}\BibitemShut {NoStop}%
\bibitem [{\citenamefont {Ramakrishnan}\ \emph {et~al.}(2024)\citenamefont
  {Ramakrishnan}, \citenamefont {Chen}, \citenamefont {Fosado}, \citenamefont
  {Tubiana}, \citenamefont {Vanderlinden}, \citenamefont {Savill},
  \citenamefont {Schnaufer},\ and\ \citenamefont
  {Michieletto}}]{Ramakrishnan2024}%
  \BibitemOpen
  \bibfield  {author} {\bibinfo {author} {\bibfnamefont {S.}~\bibnamefont
  {Ramakrishnan}}, \bibinfo {author} {\bibfnamefont {Z.}~\bibnamefont {Chen}},
  \bibinfo {author} {\bibfnamefont {Y.~A.~G.}\ \bibnamefont {Fosado}}, \bibinfo
  {author} {\bibfnamefont {L.}~\bibnamefont {Tubiana}}, \bibinfo {author}
  {\bibfnamefont {W.}~\bibnamefont {Vanderlinden}}, \bibinfo {author}
  {\bibfnamefont {N.~J.}\ \bibnamefont {Savill}}, \bibinfo {author}
  {\bibfnamefont {A.}~\bibnamefont {Schnaufer}}, \ and\ \bibinfo {author}
  {\bibfnamefont {D.}~\bibnamefont {Michieletto}},\ }\href {\doibase
  10.1103/prxlife.2.013009} {\bibfield  {journal} {\bibinfo  {journal} {PRX
  Life}\ }\textbf {\bibinfo {volume} {2}},\ \bibinfo {pages} {13009} (\bibinfo
  {year} {2024})}\BibitemShut {NoStop}%
\bibitem [{\citenamefont {Yadav}\ \emph {et~al.}(2021)\citenamefont {Yadav},
  \citenamefont {{Al Sulaiman}}, \citenamefont {Soh},\ and\ \citenamefont
  {Doyle}}]{Yadav2021}%
  \BibitemOpen
  \bibfield  {author} {\bibinfo {author} {\bibfnamefont {I.}~\bibnamefont
  {Yadav}}, \bibinfo {author} {\bibfnamefont {D.}~\bibnamefont {{Al
  Sulaiman}}}, \bibinfo {author} {\bibfnamefont {B.~W.}\ \bibnamefont {Soh}}, \
  and\ \bibinfo {author} {\bibfnamefont {P.~S.}\ \bibnamefont {Doyle}},\ }\href
  {\doibase 10.1021/acsmacrolett.1c00463} {\bibfield  {journal} {\bibinfo
  {journal} {ACS Macro Letters}\ }\textbf {\bibinfo {volume} {10}},\ \bibinfo
  {pages} {1429} (\bibinfo {year} {2021})}\BibitemShut {NoStop}%
\bibitem [{\citenamefont {Gennes}(1979)}]{Gennes1979a}%
  \BibitemOpen
  \bibfield  {author} {\bibinfo {author} {\bibfnamefont {P.~G.~D.}\
  \bibnamefont {Gennes}},\ }\href@noop {} {\emph {\bibinfo {title} {{Scaling
  concepts in polymer physics}}}}\ (\bibinfo  {publisher} {Cornell University
  Press},\ \bibinfo {year} {1979})\BibitemShut {NoStop}%
\bibitem [{\citenamefont {Perez-Morga}\ and\ \citenamefont
  {Englund}(1993)}]{Perez-Morga1993b}%
  \BibitemOpen
  \bibfield  {author} {\bibinfo {author} {\bibfnamefont {D.}~\bibnamefont
  {Perez-Morga}}\ and\ \bibinfo {author} {\bibfnamefont {P.~T.}\ \bibnamefont
  {Englund}},\ }\href {http://jcb.rupress.org/content/123/5/1069.abstract}
  {\bibfield  {journal} {\bibinfo  {journal} {J. Cell. Biol.}\ }\textbf
  {\bibinfo {volume} {123}},\ \bibinfo {pages} {1069} (\bibinfo {year}
  {1993})}\BibitemShut {NoStop}%
\bibitem [{\citenamefont {Ferguson}\ \emph {et~al.}(1994)\citenamefont
  {Ferguson}, \citenamefont {Torri}, \citenamefont {P{\'{e}}rez-Morga},
  \citenamefont {Ward},\ and\ \citenamefont {Englund}}]{Ferguson1994}%
  \BibitemOpen
  \bibfield  {author} {\bibinfo {author} {\bibfnamefont {M.~L.}\ \bibnamefont
  {Ferguson}}, \bibinfo {author} {\bibfnamefont {A.~F.}\ \bibnamefont {Torri}},
  \bibinfo {author} {\bibfnamefont {D.}~\bibnamefont {P{\'{e}}rez-Morga}},
  \bibinfo {author} {\bibfnamefont {D.~C.}\ \bibnamefont {Ward}}, \ and\
  \bibinfo {author} {\bibfnamefont {P.~T.}\ \bibnamefont {Englund}},\ }\href
  {\doibase 10.1083/jcb.126.3.631} {\bibfield  {journal} {\bibinfo  {journal}
  {Journal of Cell Biology}\ }\textbf {\bibinfo {volume} {126}},\ \bibinfo
  {pages} {631} (\bibinfo {year} {1994})}\BibitemShut {NoStop}%
\bibitem [{\citenamefont {Stuart}\ \emph {et~al.}(2005)\citenamefont {Stuart},
  \citenamefont {Schnaufer}, \citenamefont {Ernst},\ and\ \citenamefont
  {Panigrahi}}]{Stuart2005}%
  \BibitemOpen
  \bibfield  {author} {\bibinfo {author} {\bibfnamefont {K.~D.}\ \bibnamefont
  {Stuart}}, \bibinfo {author} {\bibfnamefont {A.}~\bibnamefont {Schnaufer}},
  \bibinfo {author} {\bibfnamefont {N.~L.}\ \bibnamefont {Ernst}}, \ and\
  \bibinfo {author} {\bibfnamefont {A.~K.}\ \bibnamefont {Panigrahi}},\ }\href
  {\doibase 10.1016/j.tibs.2004.12.006} {\bibfield  {journal} {\bibinfo
  {journal} {Trends in Biochemical Sciences}\ }\textbf {\bibinfo {volume}
  {30}},\ \bibinfo {pages} {97} (\bibinfo {year} {2005})}\BibitemShut {NoStop}%
\bibitem [{\citenamefont {Shapiro}\ and\ \citenamefont
  {Englund}(1995)}]{Shapiro1995}%
  \BibitemOpen
  \bibfield  {author} {\bibinfo {author} {\bibfnamefont {T.~A.}\ \bibnamefont
  {Shapiro}}\ and\ \bibinfo {author} {\bibfnamefont {P.~T.}\ \bibnamefont
  {Englund}},\ }\href
  {http://www.annualreviews.org/doi/pdf/10.1146/annurev.mi.49.100195.001001}
  {\bibfield  {journal} {\bibinfo  {journal} {Annu. Rev. Microbiol.}\ }\textbf
  {\bibinfo {volume} {49}},\ \bibinfo {pages} {117} (\bibinfo {year}
  {1995})}\BibitemShut {NoStop}%
\bibitem [{\citenamefont {Cooper}\ \emph {et~al.}(2019)\citenamefont {Cooper},
  \citenamefont {Wadsworth}, \citenamefont {Ochsenreiter}, \citenamefont
  {Ivens}, \citenamefont {Savill},\ and\ \citenamefont
  {Schnaufer}}]{Cooper2019}%
  \BibitemOpen
  \bibfield  {author} {\bibinfo {author} {\bibfnamefont {S.}~\bibnamefont
  {Cooper}}, \bibinfo {author} {\bibfnamefont {E.~S.}\ \bibnamefont
  {Wadsworth}}, \bibinfo {author} {\bibfnamefont {T.}~\bibnamefont
  {Ochsenreiter}}, \bibinfo {author} {\bibfnamefont {A.}~\bibnamefont {Ivens}},
  \bibinfo {author} {\bibfnamefont {N.~J.}\ \bibnamefont {Savill}}, \ and\
  \bibinfo {author} {\bibfnamefont {A.}~\bibnamefont {Schnaufer}},\ }\href
  {\doibase 10.1093/nar/gkz928} {\bibfield  {journal} {\bibinfo  {journal}
  {Nucleic acids research}\ }\textbf {\bibinfo {volume} {47}},\ \bibinfo
  {pages} {11304} (\bibinfo {year} {2019})}\BibitemShut {NoStop}%
\bibitem [{\citenamefont {Bernabei}\ \emph {et~al.}(2013)\citenamefont
  {Bernabei}, \citenamefont {Bacova}, \citenamefont {Moreno}, \citenamefont
  {Narros},\ and\ \citenamefont {Likos}}]{Bernabei2013}%
  \BibitemOpen
  \bibfield  {author} {\bibinfo {author} {\bibfnamefont {M.}~\bibnamefont
  {Bernabei}}, \bibinfo {author} {\bibfnamefont {P.}~\bibnamefont {Bacova}},
  \bibinfo {author} {\bibfnamefont {A.~J.}\ \bibnamefont {Moreno}}, \bibinfo
  {author} {\bibfnamefont {A.}~\bibnamefont {Narros}}, \ and\ \bibinfo {author}
  {\bibfnamefont {C.~N.}\ \bibnamefont {Likos}},\ }\href {\doibase
  10.1039/c2sm27199k} {\bibfield  {journal} {\bibinfo  {journal} {Soft Matter}\
  }\textbf {\bibinfo {volume} {9}},\ \bibinfo {pages} {1287} (\bibinfo {year}
  {2013})}\BibitemShut {NoStop}%
\bibitem [{\citenamefont {Slimani}\ \emph {et~al.}(2014)\citenamefont
  {Slimani}, \citenamefont {Bacova}, \citenamefont {Bernabei}, \citenamefont
  {Narros}, \citenamefont {Likos},\ and\ \citenamefont {Moreno}}]{Slimani2014}%
  \BibitemOpen
  \bibfield  {author} {\bibinfo {author} {\bibfnamefont {M.~Z.}\ \bibnamefont
  {Slimani}}, \bibinfo {author} {\bibfnamefont {P.}~\bibnamefont {Bacova}},
  \bibinfo {author} {\bibfnamefont {M.}~\bibnamefont {Bernabei}}, \bibinfo
  {author} {\bibfnamefont {A.}~\bibnamefont {Narros}}, \bibinfo {author}
  {\bibfnamefont {C.~N.}\ \bibnamefont {Likos}}, \ and\ \bibinfo {author}
  {\bibfnamefont {A.~J.}\ \bibnamefont {Moreno}},\ }\href {\doibase
  10.1021/mz500117v} {\bibfield  {journal} {\bibinfo  {journal} {ACS Macro
  Letters}\ }\textbf {\bibinfo {volume} {3}},\ \bibinfo {pages} {611} (\bibinfo
  {year} {2014})},\ \Eprint {http://arxiv.org/abs/1309.1061} {arXiv:1309.1061}
  \BibitemShut {NoStop}%
\bibitem [{\citenamefont {Staňo}\ \emph {et~al.}(2022)\citenamefont {Staňo},
  \citenamefont {Likos},\ and\ \citenamefont {Smrek}}]{Stano2022}%
  \BibitemOpen
  \bibfield  {author} {\bibinfo {author} {\bibfnamefont {R.}~\bibnamefont
  {Staňo}}, \bibinfo {author} {\bibfnamefont {C.~N.}\ \bibnamefont {Likos}}, \
  and\ \bibinfo {author} {\bibfnamefont {J.}~\bibnamefont {Smrek}},\ }\href
  {\doibase 10.1039/d2sm01177h} {\bibfield  {journal} {\bibinfo  {journal}
  {Soft Matter}\ }\textbf {\bibinfo {volume} {19}},\ \bibinfo {pages} {17}
  (\bibinfo {year} {2022})}\BibitemShut {NoStop}%
\bibitem [{\citenamefont {Silver}\ \emph {et~al.}(1986)\citenamefont {Silver},
  \citenamefont {Torri},\ and\ \citenamefont {Hajduk}}]{Silver1986}%
  \BibitemOpen
  \bibfield  {author} {\bibinfo {author} {\bibfnamefont {L.~E.}\ \bibnamefont
  {Silver}}, \bibinfo {author} {\bibfnamefont {A.~F.}\ \bibnamefont {Torri}}, \
  and\ \bibinfo {author} {\bibfnamefont {S.~L.}\ \bibnamefont {Hajduk}},\
  }\href {http://www.ncbi.nlm.nih.gov/pubmed/3022936} {\bibfield  {journal}
  {\bibinfo  {journal} {Cell}\ }\textbf {\bibinfo {volume} {47}},\ \bibinfo
  {pages} {537} (\bibinfo {year} {1986})}\BibitemShut {NoStop}%
\bibitem [{\citenamefont {Tubiana}\ \emph {et~al.}(2024)\citenamefont
  {Tubiana}, \citenamefont {Alexander}, \citenamefont {Barbensi}, \citenamefont
  {Buck}, \citenamefont {Cartwright}, \citenamefont {Chwastyk}, \citenamefont
  {Cieplak}, \citenamefont {Coluzza}, \citenamefont {{\v{C}}opar},
  \citenamefont {Craik}, \citenamefont {{Di Stefano}}, \citenamefont
  {Everaers}, \citenamefont {Fa{\'{i}}sca}, \citenamefont {Ferrari},
  \citenamefont {Giacometti}, \citenamefont {Goundaroulis}, \citenamefont
  {Haglund}, \citenamefont {Hou}, \citenamefont {Ilieva}, \citenamefont
  {Jackson}, \citenamefont {Japaridze}, \citenamefont {Kaplan}, \citenamefont
  {Klotz}, \citenamefont {Li}, \citenamefont {Likos}, \citenamefont
  {Locatelli}, \citenamefont {L{\'{o}}pez-Le{\'{o}}n}, \citenamefont {Machon},
  \citenamefont {Micheletti}, \citenamefont {Michieletto}, \citenamefont
  {Niemi}, \citenamefont {Niemyska}, \citenamefont {Niewieczerzal},
  \citenamefont {Nitti}, \citenamefont {Orlandini}, \citenamefont {Pasquali},
  \citenamefont {Perlinska}, \citenamefont {Podgornik}, \citenamefont
  {Potestio}, \citenamefont {Pugno}, \citenamefont {Ravnik}, \citenamefont
  {Ricca}, \citenamefont {Rohwer}, \citenamefont {Rosa}, \citenamefont {Smrek},
  \citenamefont {Souslov}, \citenamefont {Stasiak}, \citenamefont {Steer},
  \citenamefont {Su{\l}kowska}, \citenamefont {Su{\l}kowski}, \citenamefont
  {Sumners}, \citenamefont {Svaneborg}, \citenamefont {Szymczak}, \citenamefont
  {Tarenzi}, \citenamefont {Travasso}, \citenamefont {Virnau}, \citenamefont
  {Vlassopoulos}, \citenamefont {Ziherl},\ and\ \citenamefont
  {{\v{Z}}umer}}]{Tubiana2024}%
  \BibitemOpen
  \bibfield  {author} {\bibinfo {author} {\bibfnamefont {L.}~\bibnamefont
  {Tubiana}}, \bibinfo {author} {\bibfnamefont {G.~P.}\ \bibnamefont
  {Alexander}}, \bibinfo {author} {\bibfnamefont {A.}~\bibnamefont {Barbensi}},
  \bibinfo {author} {\bibfnamefont {D.}~\bibnamefont {Buck}}, \bibinfo {author}
  {\bibfnamefont {J.~H.}\ \bibnamefont {Cartwright}}, \bibinfo {author}
  {\bibfnamefont {M.}~\bibnamefont {Chwastyk}}, \bibinfo {author}
  {\bibfnamefont {M.}~\bibnamefont {Cieplak}}, \bibinfo {author} {\bibfnamefont
  {I.}~\bibnamefont {Coluzza}}, \bibinfo {author} {\bibfnamefont
  {S.}~\bibnamefont {{\v{C}}opar}}, \bibinfo {author} {\bibfnamefont {D.~J.}\
  \bibnamefont {Craik}}, \bibinfo {author} {\bibfnamefont {M.}~\bibnamefont
  {{Di Stefano}}}, \bibinfo {author} {\bibfnamefont {R.}~\bibnamefont
  {Everaers}}, \bibinfo {author} {\bibfnamefont {P.~F.}\ \bibnamefont
  {Fa{\'{i}}sca}}, \bibinfo {author} {\bibfnamefont {F.}~\bibnamefont
  {Ferrari}}, \bibinfo {author} {\bibfnamefont {A.}~\bibnamefont {Giacometti}},
  \bibinfo {author} {\bibfnamefont {D.}~\bibnamefont {Goundaroulis}}, \bibinfo
  {author} {\bibfnamefont {E.}~\bibnamefont {Haglund}}, \bibinfo {author}
  {\bibfnamefont {Y.~M.}\ \bibnamefont {Hou}}, \bibinfo {author} {\bibfnamefont
  {N.}~\bibnamefont {Ilieva}}, \bibinfo {author} {\bibfnamefont {S.~E.}\
  \bibnamefont {Jackson}}, \bibinfo {author} {\bibfnamefont {A.}~\bibnamefont
  {Japaridze}}, \bibinfo {author} {\bibfnamefont {N.}~\bibnamefont {Kaplan}},
  \bibinfo {author} {\bibfnamefont {A.~R.}\ \bibnamefont {Klotz}}, \bibinfo
  {author} {\bibfnamefont {H.}~\bibnamefont {Li}}, \bibinfo {author}
  {\bibfnamefont {C.~N.}\ \bibnamefont {Likos}}, \bibinfo {author}
  {\bibfnamefont {E.}~\bibnamefont {Locatelli}}, \bibinfo {author}
  {\bibfnamefont {T.}~\bibnamefont {L{\'{o}}pez-Le{\'{o}}n}}, \bibinfo {author}
  {\bibfnamefont {T.}~\bibnamefont {Machon}}, \bibinfo {author} {\bibfnamefont
  {C.}~\bibnamefont {Micheletti}}, \bibinfo {author} {\bibfnamefont
  {D.}~\bibnamefont {Michieletto}}, \bibinfo {author} {\bibfnamefont
  {A.}~\bibnamefont {Niemi}}, \bibinfo {author} {\bibfnamefont
  {W.}~\bibnamefont {Niemyska}}, \bibinfo {author} {\bibfnamefont
  {S.}~\bibnamefont {Niewieczerzal}}, \bibinfo {author} {\bibfnamefont
  {F.}~\bibnamefont {Nitti}}, \bibinfo {author} {\bibfnamefont
  {E.}~\bibnamefont {Orlandini}}, \bibinfo {author} {\bibfnamefont
  {S.}~\bibnamefont {Pasquali}}, \bibinfo {author} {\bibfnamefont {A.~P.}\
  \bibnamefont {Perlinska}}, \bibinfo {author} {\bibfnamefont {R.}~\bibnamefont
  {Podgornik}}, \bibinfo {author} {\bibfnamefont {R.}~\bibnamefont {Potestio}},
  \bibinfo {author} {\bibfnamefont {N.~M.}\ \bibnamefont {Pugno}}, \bibinfo
  {author} {\bibfnamefont {M.}~\bibnamefont {Ravnik}}, \bibinfo {author}
  {\bibfnamefont {R.}~\bibnamefont {Ricca}}, \bibinfo {author} {\bibfnamefont
  {C.~M.}\ \bibnamefont {Rohwer}}, \bibinfo {author} {\bibfnamefont
  {A.}~\bibnamefont {Rosa}}, \bibinfo {author} {\bibfnamefont {J.}~\bibnamefont
  {Smrek}}, \bibinfo {author} {\bibfnamefont {A.}~\bibnamefont {Souslov}},
  \bibinfo {author} {\bibfnamefont {A.}~\bibnamefont {Stasiak}}, \bibinfo
  {author} {\bibfnamefont {D.}~\bibnamefont {Steer}}, \bibinfo {author}
  {\bibfnamefont {J.}~\bibnamefont {Su{\l}kowska}}, \bibinfo {author}
  {\bibfnamefont {P.}~\bibnamefont {Su{\l}kowski}}, \bibinfo {author}
  {\bibfnamefont {D.~W.~L.}\ \bibnamefont {Sumners}}, \bibinfo {author}
  {\bibfnamefont {C.}~\bibnamefont {Svaneborg}}, \bibinfo {author}
  {\bibfnamefont {P.}~\bibnamefont {Szymczak}}, \bibinfo {author}
  {\bibfnamefont {T.}~\bibnamefont {Tarenzi}}, \bibinfo {author} {\bibfnamefont
  {R.}~\bibnamefont {Travasso}}, \bibinfo {author} {\bibfnamefont
  {P.}~\bibnamefont {Virnau}}, \bibinfo {author} {\bibfnamefont
  {D.}~\bibnamefont {Vlassopoulos}}, \bibinfo {author} {\bibfnamefont
  {P.}~\bibnamefont {Ziherl}}, \ and\ \bibinfo {author} {\bibfnamefont
  {S.}~\bibnamefont {{\v{Z}}umer}},\ }\href
  {https://doi.org/10.1016/j.physrep.2024.04.002} {\bibfield  {journal}
  {\bibinfo  {journal} {Physics Reports}\ }\textbf {\bibinfo {volume} {1075}},\
  \bibinfo {pages} {1} (\bibinfo {year} {2024})}\BibitemShut {NoStop}%
\bibitem [{\citenamefont {Bates}\ and\ \citenamefont
  {Maxwell}(2005)}]{Bates2005}%
  \BibitemOpen
  \bibfield  {author} {\bibinfo {author} {\bibfnamefont {A.}~\bibnamefont
  {Bates}}\ and\ \bibinfo {author} {\bibfnamefont {A.}~\bibnamefont
  {Maxwell}},\ }\href@noop {} {\emph {\bibinfo {title} {{DNA topology}}}}\
  (\bibinfo  {publisher} {Oxford University Press},\ \bibinfo {year}
  {2005})\BibitemShut {NoStop}%
\bibitem [{\citenamefont {Gamill}\ \emph {et~al.}(2024)\citenamefont {Gamill},
  \citenamefont {Holmes}, \citenamefont {Provan}, \citenamefont {Wiggins},
  \citenamefont {Ruskov{\'{a}}}, \citenamefont {Whittle}, \citenamefont
  {Catley}, \citenamefont {Main}, \citenamefont {Shephard}, \citenamefont
  {Bryant}, \citenamefont {Ra{\v{c}}ko}, \citenamefont {Colloms},\ and\
  \citenamefont {Pyne}}]{Gamill2024}%
  \BibitemOpen
  \bibfield  {author} {\bibinfo {author} {\bibfnamefont {M.~C.}\ \bibnamefont
  {Gamill}}, \bibinfo {author} {\bibfnamefont {E.~P.}\ \bibnamefont {Holmes}},
  \bibinfo {author} {\bibfnamefont {J.~I.}\ \bibnamefont {Provan}}, \bibinfo
  {author} {\bibfnamefont {L.}~\bibnamefont {Wiggins}}, \bibinfo {author}
  {\bibfnamefont {R.}~\bibnamefont {Ruskov{\'{a}}}}, \bibinfo {author}
  {\bibfnamefont {S.}~\bibnamefont {Whittle}}, \bibinfo {author} {\bibfnamefont
  {T.~E.}\ \bibnamefont {Catley}}, \bibinfo {author} {\bibfnamefont {K.~H.~S.}\
  \bibnamefont {Main}}, \bibinfo {author} {\bibfnamefont {N.}~\bibnamefont
  {Shephard}}, \bibinfo {author} {\bibfnamefont {H.~E.}\ \bibnamefont
  {Bryant}}, \bibinfo {author} {\bibfnamefont {D.}~\bibnamefont {Ra{\v{c}}ko}},
  \bibinfo {author} {\bibfnamefont {S.~D.}\ \bibnamefont {Colloms}}, \ and\
  \bibinfo {author} {\bibfnamefont {A.~L.~B.}\ \bibnamefont {Pyne}},\ }\href
  {http://biorxiv.org/lookup/doi/10.1101/2024.06.28.601212} {\bibfield
  {journal} {\bibinfo  {journal} {bioRxiv}\ ,\ \bibinfo {pages} {1}} (\bibinfo
  {year} {2024})}\BibitemShut {NoStop}%
\bibitem [{\citenamefont {Chen}\ \emph
  {et~al.}(1995{\natexlab{a}})\citenamefont {Chen}, \citenamefont {Rauch},
  \citenamefont {White}, \citenamefont {Englund},\ and\ \citenamefont
  {Cozzarelli}}]{Chen1995}%
  \BibitemOpen
  \bibfield  {author} {\bibinfo {author} {\bibfnamefont {J.}~\bibnamefont
  {Chen}}, \bibinfo {author} {\bibfnamefont {C.~A.}\ \bibnamefont {Rauch}},
  \bibinfo {author} {\bibfnamefont {J.~H.}\ \bibnamefont {White}}, \bibinfo
  {author} {\bibfnamefont {P.~T.}\ \bibnamefont {Englund}}, \ and\ \bibinfo
  {author} {\bibfnamefont {N.}~\bibnamefont {Cozzarelli}},\ }\href
  {http://www.ncbi.nlm.nih.gov/pubmed/7813018} {\bibfield  {journal} {\bibinfo
  {journal} {Cell}\ }\textbf {\bibinfo {volume} {80}},\ \bibinfo {pages} {61}
  (\bibinfo {year} {1995}{\natexlab{a}})}\BibitemShut {NoStop}%
\bibitem [{\citenamefont {Ibrahim}\ \emph {et~al.}(2019)\citenamefont
  {Ibrahim}, \citenamefont {Liu}, \citenamefont {Klingbeil}, \citenamefont
  {Diao},\ and\ \citenamefont {Arsuaga}}]{Ibrahim2019}%
  \BibitemOpen
  \bibfield  {author} {\bibinfo {author} {\bibfnamefont {L.}~\bibnamefont
  {Ibrahim}}, \bibinfo {author} {\bibfnamefont {P.}~\bibnamefont {Liu}},
  \bibinfo {author} {\bibfnamefont {M.}~\bibnamefont {Klingbeil}}, \bibinfo
  {author} {\bibfnamefont {Y.}~\bibnamefont {Diao}}, \ and\ \bibinfo {author}
  {\bibfnamefont {J.}~\bibnamefont {Arsuaga}},\ }\href {\doibase
  10.1088/1751-8121/aaf15f} {\bibfield  {journal} {\bibinfo  {journal} {Journal
  of Physics A: Mathematical and Theoretical}\ }\textbf {\bibinfo {volume}
  {52}} (\bibinfo {year} {2019}),\ 10.1088/1751-8121/aaf15f}\BibitemShut
  {NoStop}%
\bibitem [{\citenamefont {Chen}\ \emph
  {et~al.}(1995{\natexlab{b}})\citenamefont {Chen}, \citenamefont {Englund},\
  and\ \citenamefont {Cozzarelli}}]{Chen1995a}%
  \BibitemOpen
  \bibfield  {author} {\bibinfo {author} {\bibfnamefont {J.}~\bibnamefont
  {Chen}}, \bibinfo {author} {\bibfnamefont {P.~T.}\ \bibnamefont {Englund}}, \
  and\ \bibinfo {author} {\bibfnamefont {N.~R.}\ \bibnamefont {Cozzarelli}},\
  }\href {http://www.ncbi.nlm.nih.gov/pmc/articles/PMC394759/} {\bibfield
  {journal} {\bibinfo  {journal} {EMBO J.}\ }\textbf {\bibinfo {volume} {14}},\
  \bibinfo {pages} {6339} (\bibinfo {year} {1995}{\natexlab{b}})}\BibitemShut
  {NoStop}%
\bibitem [{\citenamefont {Michieletto}\ \emph {et~al.}(2015)\citenamefont
  {Michieletto}, \citenamefont {Marenduzzo},\ and\ \citenamefont
  {Orlandini}}]{Michieletto2014a}%
  \BibitemOpen
  \bibfield  {author} {\bibinfo {author} {\bibfnamefont {D.}~\bibnamefont
  {Michieletto}}, \bibinfo {author} {\bibfnamefont {D.}~\bibnamefont
  {Marenduzzo}}, \ and\ \bibinfo {author} {\bibfnamefont {E.}~\bibnamefont
  {Orlandini}},\ }\href
  {http://iopscience.iop.org/article/10.1088/1478-3975/12/3/036001} {\bibfield
  {journal} {\bibinfo  {journal} {Phys. Biol.}\ }\textbf {\bibinfo {volume}
  {12}},\ \bibinfo {pages} {036001} (\bibinfo {year} {2015})}\BibitemShut
  {NoStop}%
\bibitem [{\citenamefont {Bonhivers}\ \emph {et~al.}(2008)\citenamefont
  {Bonhivers}, \citenamefont {Landrein}, \citenamefont {Decossas},\ and\
  \citenamefont {Robinson}}]{Bonhivers2008}%
  \BibitemOpen
  \bibfield  {author} {\bibinfo {author} {\bibfnamefont {M.}~\bibnamefont
  {Bonhivers}}, \bibinfo {author} {\bibfnamefont {N.}~\bibnamefont {Landrein}},
  \bibinfo {author} {\bibfnamefont {M.}~\bibnamefont {Decossas}}, \ and\
  \bibinfo {author} {\bibfnamefont {D.~R.}\ \bibnamefont {Robinson}},\ }\href
  {\doibase 10.1186/1756-3305-1-21} {\bibfield  {journal} {\bibinfo  {journal}
  {Parasites {\&} vectors}\ }\textbf {\bibinfo {volume} {1}},\ \bibinfo {pages}
  {21} (\bibinfo {year} {2008})}\BibitemShut {NoStop}%
\bibitem [{\citenamefont {Amodeo}\ \emph {et~al.}(2021)\citenamefont {Amodeo},
  \citenamefont {Kalichava}, \citenamefont {Fradera-Sola}, \citenamefont
  {Bertiaux-Lequoy}, \citenamefont {Guichard}, \citenamefont {Butter},\ and\
  \citenamefont {Ochsenreiter}}]{Amodeo2021}%
  \BibitemOpen
  \bibfield  {author} {\bibinfo {author} {\bibfnamefont {S.}~\bibnamefont
  {Amodeo}}, \bibinfo {author} {\bibfnamefont {A.}~\bibnamefont {Kalichava}},
  \bibinfo {author} {\bibfnamefont {A.}~\bibnamefont {Fradera-Sola}}, \bibinfo
  {author} {\bibfnamefont {E.}~\bibnamefont {Bertiaux-Lequoy}}, \bibinfo
  {author} {\bibfnamefont {P.}~\bibnamefont {Guichard}}, \bibinfo {author}
  {\bibfnamefont {F.}~\bibnamefont {Butter}}, \ and\ \bibinfo {author}
  {\bibfnamefont {T.}~\bibnamefont {Ochsenreiter}},\ }\href {\doibase
  10.1242/jcs.254300} {\bibfield  {journal} {\bibinfo  {journal} {Journal of
  Cell Science}\ }\textbf {\bibinfo {volume} {134}} (\bibinfo {year} {2021}),\
  10.1242/jcs.254300}\BibitemShut {NoStop}%
\bibitem [{\citenamefont {Kalichava}\ and\ \citenamefont
  {Ochsenreiter}(2021)}]{Kalichava2021}%
  \BibitemOpen
  \bibfield  {author} {\bibinfo {author} {\bibfnamefont {A.}~\bibnamefont
  {Kalichava}}\ and\ \bibinfo {author} {\bibfnamefont {T.}~\bibnamefont
  {Ochsenreiter}},\ }\href {\doibase 10.1098/rsob.210132} {\bibfield  {journal}
  {\bibinfo  {journal} {Open Biology}\ }\textbf {\bibinfo {volume} {11}},\
  \bibinfo {pages} {14} (\bibinfo {year} {2021})}\BibitemShut {NoStop}%
\bibitem [{\citenamefont {Hoffmann}\ \emph {et~al.}(2018)\citenamefont
  {Hoffmann}, \citenamefont {K{\"{a}}ser}, \citenamefont {Jakob}, \citenamefont
  {Amodeo}, \citenamefont {Peitsch}, \citenamefont {T{\'{y}}c}, \citenamefont
  {Vaughan}, \citenamefont {Zuber}, \citenamefont {Schneider},\ and\
  \citenamefont {Ochsenreiter}}]{Hoffmann2018}%
  \BibitemOpen
  \bibfield  {author} {\bibinfo {author} {\bibfnamefont {A.}~\bibnamefont
  {Hoffmann}}, \bibinfo {author} {\bibfnamefont {S.}~\bibnamefont
  {K{\"{a}}ser}}, \bibinfo {author} {\bibfnamefont {M.}~\bibnamefont {Jakob}},
  \bibinfo {author} {\bibfnamefont {S.}~\bibnamefont {Amodeo}}, \bibinfo
  {author} {\bibfnamefont {C.}~\bibnamefont {Peitsch}}, \bibinfo {author}
  {\bibfnamefont {J.}~\bibnamefont {T{\'{y}}c}}, \bibinfo {author}
  {\bibfnamefont {S.}~\bibnamefont {Vaughan}}, \bibinfo {author} {\bibfnamefont
  {B.}~\bibnamefont {Zuber}}, \bibinfo {author} {\bibfnamefont
  {A.}~\bibnamefont {Schneider}}, \ and\ \bibinfo {author} {\bibfnamefont
  {T.}~\bibnamefont {Ochsenreiter}},\ }\href {\doibase 10.1073/pnas.1716582115}
  {\bibfield  {journal} {\bibinfo  {journal} {Proceedings of the National
  Academy of Sciences of the United States of America}\ }\textbf {\bibinfo
  {volume} {115}},\ \bibinfo {pages} {E1809} (\bibinfo {year}
  {2018})}\BibitemShut {NoStop}%
\bibitem [{\citenamefont {Schnaufer}(2010)}]{Schnaufer2010}%
  \BibitemOpen
  \bibfield  {author} {\bibinfo {author} {\bibfnamefont {A.}~\bibnamefont
  {Schnaufer}},\ }\href {\doibase 10.1016/j.pt.2010.08.001.Evolution}
  {\bibfield  {journal} {\bibinfo  {journal} {Trends Parasitol.}\ }\textbf
  {\bibinfo {volume} {26}},\ \bibinfo {pages} {557} (\bibinfo {year}
  {2010})}\BibitemShut {NoStop}%
\bibitem [{Note1()}]{Note1}%
  \BibitemOpen
  \bibinfo {note} {To my knowledge this hypothesis was first raised and
  discussed in a series of exchanges between the author, Luca Tubiana and Achim
  Schnaufer.}\BibitemShut {Stop}%
\bibitem [{\citenamefont {Soh}\ and\ \citenamefont {Doyle}(2021)}]{Soh2021}%
  \BibitemOpen
  \bibfield  {author} {\bibinfo {author} {\bibfnamefont {B.~W.}\ \bibnamefont
  {Soh}}\ and\ \bibinfo {author} {\bibfnamefont {P.~S.}\ \bibnamefont
  {Doyle}},\ }\href {\doibase 10.1021/acsmacrolett.1c00299} {\bibfield
  {journal} {\bibinfo  {journal} {ACS Macro Letters}\ }\textbf {\bibinfo
  {volume} {10}},\ \bibinfo {pages} {880} (\bibinfo {year} {2021})}\BibitemShut
  {NoStop}%
\bibitem [{\citenamefont {Polson}\ \emph {et~al.}(2021)\citenamefont {Polson},
  \citenamefont {Garcia},\ and\ \citenamefont {Klotz}}]{Polson2021}%
  \BibitemOpen
  \bibfield  {author} {\bibinfo {author} {\bibfnamefont {J.~M.}\ \bibnamefont
  {Polson}}, \bibinfo {author} {\bibfnamefont {E.~J.}\ \bibnamefont {Garcia}},
  \ and\ \bibinfo {author} {\bibfnamefont {A.~R.}\ \bibnamefont {Klotz}},\
  }\href@noop {} {\bibfield  {journal} {\bibinfo  {journal} {Soft Matter}\
  }\textbf {\bibinfo {volume} {17}},\ \bibinfo {pages} {10505} (\bibinfo {year}
  {2021})}\BibitemShut {NoStop}%
\bibitem [{\citenamefont {Ragotskie}\ \emph {et~al.}(2024)\citenamefont
  {Ragotskie}, \citenamefont {Morrison}, \citenamefont {Stackhouse},
  \citenamefont {Blair},\ and\ \citenamefont {Klotz}}]{Ragotskie2024}%
  \BibitemOpen
  \bibfield  {author} {\bibinfo {author} {\bibfnamefont {J.}~\bibnamefont
  {Ragotskie}}, \bibinfo {author} {\bibfnamefont {N.}~\bibnamefont {Morrison}},
  \bibinfo {author} {\bibfnamefont {C.}~\bibnamefont {Stackhouse}}, \bibinfo
  {author} {\bibfnamefont {R.~C.}\ \bibnamefont {Blair}}, \ and\ \bibinfo
  {author} {\bibfnamefont {A.~R.}\ \bibnamefont {Klotz}},\ }\href
  {https://onlinelibrary.wiley.com/doi/abs/10.1002/pol.20230392} {\bibfield
  {journal} {\bibinfo  {journal} {Journal of Polymer Science}\ }\textbf
  {\bibinfo {volume} {62}},\ \bibinfo {pages} {1287} (\bibinfo {year}
  {2024})}\BibitemShut {NoStop}%
\bibitem [{\citenamefont {Tubiana}\ \emph {et~al.}(2022)\citenamefont
  {Tubiana}, \citenamefont {Ferrari},\ and\ \citenamefont
  {Orlandini}}]{Tubiana2022}%
  \BibitemOpen
  \bibfield  {author} {\bibinfo {author} {\bibfnamefont {L.}~\bibnamefont
  {Tubiana}}, \bibinfo {author} {\bibfnamefont {F.}~\bibnamefont {Ferrari}}, \
  and\ \bibinfo {author} {\bibfnamefont {E.}~\bibnamefont {Orlandini}},\ }\href
  {https://doi.org/10.1103/PhysRevLett.129.227801} {\bibfield  {journal}
  {\bibinfo  {journal} {Physical Review Letters}\ }\textbf {\bibinfo {volume}
  {129}},\ \bibinfo {pages} {227801} (\bibinfo {year} {2022})}\BibitemShut
  {NoStop}%
\bibitem [{\citenamefont {Klotz}\ \emph {et~al.}()\citenamefont {Klotz},
  \citenamefont {Anderson},\ and\ \citenamefont
  {Dimitriyev}}]{Klotz2024chirality}%
  \BibitemOpen
  \bibfield  {author} {\bibinfo {author} {\bibfnamefont {A.~R.}\ \bibnamefont
  {Klotz}}, \bibinfo {author} {\bibfnamefont {C.~J.}\ \bibnamefont {Anderson}},
  \ and\ \bibinfo {author} {\bibfnamefont {M.~S.}\ \bibnamefont {Dimitriyev}},\
  }\href@noop {} {\bibfield  {journal} {\bibinfo  {journal} {arxiv}\ }}\Eprint
  {http://arxiv.org/abs/arXiv:2406.13590v1} {arXiv:arXiv:2406.13590v1}
  \BibitemShut {NoStop}%
\bibitem [{\citenamefont {Shankar}\ and\ \citenamefont
  {Nelson}(2021)}]{Shankar2021}%
  \BibitemOpen
  \bibfield  {author} {\bibinfo {author} {\bibfnamefont {S.}~\bibnamefont
  {Shankar}}\ and\ \bibinfo {author} {\bibfnamefont {D.~R.}\ \bibnamefont
  {Nelson}},\ }\href@noop {} {\bibfield  {journal} {\bibinfo  {journal}
  {Physical Review E}\ }\textbf {\bibinfo {volume} {104}},\ \bibinfo {pages}
  {1} (\bibinfo {year} {2021})}\BibitemShut {NoStop}%
\bibitem [{\citenamefont {Soh}\ \emph {et~al.}(2020)\citenamefont {Soh},
  \citenamefont {Khorshid}, \citenamefont {{Al Sulaiman}},\ and\ \citenamefont
  {Doyle}}]{Soh2020}%
  \BibitemOpen
  \bibfield  {author} {\bibinfo {author} {\bibfnamefont {B.~W.}\ \bibnamefont
  {Soh}}, \bibinfo {author} {\bibfnamefont {A.}~\bibnamefont {Khorshid}},
  \bibinfo {author} {\bibfnamefont {D.}~\bibnamefont {{Al Sulaiman}}}, \ and\
  \bibinfo {author} {\bibfnamefont {P.~S.}\ \bibnamefont {Doyle}},\ }\href
  {\doibase 10.1021/acs.macromol.0c01706} {\bibfield  {journal} {\bibinfo
  {journal} {Macromolecules}\ }\textbf {\bibinfo {volume} {53}},\ \bibinfo
  {pages} {8502} (\bibinfo {year} {2020})}\BibitemShut {NoStop}%
\bibitem [{\citenamefont {Raphael}\ \emph {et~al.}(1997)\citenamefont
  {Raphael}, \citenamefont {Gay},\ and\ \citenamefont
  {de~Gennes}}]{Raphael1997}%
  \BibitemOpen
  \bibfield  {author} {\bibinfo {author} {\bibfnamefont {E.}~\bibnamefont
  {Raphael}}, \bibinfo {author} {\bibfnamefont {C.}~\bibnamefont {Gay}}, \ and\
  \bibinfo {author} {\bibfnamefont {P.~G.}\ \bibnamefont {de~Gennes}},\ }\href
  {http://link.springer.com/article/10.1007/BF02770756} {\bibfield  {journal}
  {\bibinfo  {journal} {J. Stat. Phys.}\ }\textbf {\bibinfo {volume} {89}},\
  \bibinfo {pages} {111} (\bibinfo {year} {1997})}\BibitemShut {NoStop}%
\bibitem [{\citenamefont {Pickett}(2006)}]{Pickett2006}%
  \BibitemOpen
  \bibfield  {author} {\bibinfo {author} {\bibfnamefont {G.~T.}\ \bibnamefont
  {Pickett}},\ }\href {\doibase 10.1209/epl/i2006-10302-7} {\bibfield
  {journal} {\bibinfo  {journal} {EPL}\ }\textbf {\bibinfo {volume} {76}},\
  \bibinfo {pages} {616} (\bibinfo {year} {2006})}\BibitemShut {NoStop}%
\bibitem [{\citenamefont {Speed}\ \emph {et~al.}(2024)\citenamefont {Speed},
  \citenamefont {Atabay}, \citenamefont {Peng}, \citenamefont {Gupta},
  \citenamefont {Sommer}, \citenamefont {Lang},\ and\ \citenamefont
  {Krieg}}]{Speed2024}%
  \BibitemOpen
  \bibfield  {author} {\bibinfo {author} {\bibfnamefont {S.}~\bibnamefont
  {Speed}}, \bibinfo {author} {\bibfnamefont {A.}~\bibnamefont {Atabay}},
  \bibinfo {author} {\bibfnamefont {Y.-h.}\ \bibnamefont {Peng}}, \bibinfo
  {author} {\bibfnamefont {K.}~\bibnamefont {Gupta}}, \bibinfo {author}
  {\bibfnamefont {J.-u.}\ \bibnamefont {Sommer}}, \bibinfo {author}
  {\bibfnamefont {M.}~\bibnamefont {Lang}}, \ and\ \bibinfo {author}
  {\bibfnamefont {E.}~\bibnamefont {Krieg}},\ }\href {\doibase
  doi.org/10.1101/2024.07.12.603212} {\bibfield  {journal} {\bibinfo  {journal}
  {bioRxiv}\ } (\bibinfo {year} {2024}),\ doi.org/10.1101/2024.07.12.603212},\
  \Eprint {http://arxiv.org/abs/2024.07.12.603212} {arXiv:2024.07.12.603212}
  \BibitemShut {NoStop}%
\bibitem [{\citenamefont {Palombo}\ \emph {et~al.}(2023)\citenamefont
  {Palombo}, \citenamefont {Weir}, \citenamefont {Michieletto},\ and\
  \citenamefont {Fosado}}]{Palombo2023}%
  \BibitemOpen
  \bibfield  {author} {\bibinfo {author} {\bibfnamefont {G.}~\bibnamefont
  {Palombo}}, \bibinfo {author} {\bibfnamefont {S.}~\bibnamefont {Weir}},
  \bibinfo {author} {\bibfnamefont {D.}~\bibnamefont {Michieletto}}, \ and\
  \bibinfo {author} {\bibfnamefont {Y.~A.~G.}\ \bibnamefont {Fosado}},\ }\href
  {http://arxiv.org/abs/2308.09689} {\bibfield  {journal} {\bibinfo  {journal}
  {arXiv:2308.09689v1}\ } (\bibinfo {year} {2023})},\ \Eprint
  {http://arxiv.org/abs/2308.09689} {arXiv:2308.09689} \BibitemShut {NoStop}%
\bibitem [{\citenamefont {Datta}\ \emph {et~al.}(2020)\citenamefont {Datta},
  \citenamefont {Kato}, \citenamefont {Higashiharaguchi}, \citenamefont
  {Aratsu}, \citenamefont {Isobe}, \citenamefont {Saito}, \citenamefont
  {Prabhu}, \citenamefont {Kitamoto}, \citenamefont {Hollamby}, \citenamefont
  {Smith}, \citenamefont {Dagleish}, \citenamefont {Mahmoudi}, \citenamefont
  {Pesce}, \citenamefont {Perego}, \citenamefont {Pavan},\ and\ \citenamefont
  {Yagai}}]{Datta2020}%
  \BibitemOpen
  \bibfield  {author} {\bibinfo {author} {\bibfnamefont {S.}~\bibnamefont
  {Datta}}, \bibinfo {author} {\bibfnamefont {Y.}~\bibnamefont {Kato}},
  \bibinfo {author} {\bibfnamefont {S.}~\bibnamefont {Higashiharaguchi}},
  \bibinfo {author} {\bibfnamefont {K.}~\bibnamefont {Aratsu}}, \bibinfo
  {author} {\bibfnamefont {A.}~\bibnamefont {Isobe}}, \bibinfo {author}
  {\bibfnamefont {T.}~\bibnamefont {Saito}}, \bibinfo {author} {\bibfnamefont
  {D.~D.}\ \bibnamefont {Prabhu}}, \bibinfo {author} {\bibfnamefont
  {Y.}~\bibnamefont {Kitamoto}}, \bibinfo {author} {\bibfnamefont {M.~J.}\
  \bibnamefont {Hollamby}}, \bibinfo {author} {\bibfnamefont {A.~J.}\
  \bibnamefont {Smith}}, \bibinfo {author} {\bibfnamefont {R.}~\bibnamefont
  {Dagleish}}, \bibinfo {author} {\bibfnamefont {N.}~\bibnamefont {Mahmoudi}},
  \bibinfo {author} {\bibfnamefont {L.}~\bibnamefont {Pesce}}, \bibinfo
  {author} {\bibfnamefont {C.}~\bibnamefont {Perego}}, \bibinfo {author}
  {\bibfnamefont {G.~M.}\ \bibnamefont {Pavan}}, \ and\ \bibinfo {author}
  {\bibfnamefont {S.}~\bibnamefont {Yagai}},\ }\href {\doibase
  10.1038/s41586-020-2445-z} {\bibfield  {journal} {\bibinfo  {journal}
  {Nature}\ }\textbf {\bibinfo {volume} {583}},\ \bibinfo {pages} {400}
  (\bibinfo {year} {2020})}\BibitemShut {NoStop}%
\bibitem [{\citenamefont {August}\ \emph {et~al.}(2020)\citenamefont {August},
  \citenamefont {Dryfe}, \citenamefont {Haigh}, \citenamefont {Kent},
  \citenamefont {Leigh}, \citenamefont {Lemonnier}, \citenamefont {Li},
  \citenamefont {Muryn}, \citenamefont {Palmer}, \citenamefont {Song},
  \citenamefont {Whitehead},\ and\ \citenamefont {Young}}]{August2020}%
  \BibitemOpen
  \bibfield  {author} {\bibinfo {author} {\bibfnamefont {D.~P.}\ \bibnamefont
  {August}}, \bibinfo {author} {\bibfnamefont {R.~A.}\ \bibnamefont {Dryfe}},
  \bibinfo {author} {\bibfnamefont {S.~J.}\ \bibnamefont {Haigh}}, \bibinfo
  {author} {\bibfnamefont {P.~R.}\ \bibnamefont {Kent}}, \bibinfo {author}
  {\bibfnamefont {D.~A.}\ \bibnamefont {Leigh}}, \bibinfo {author}
  {\bibfnamefont {J.~F.}\ \bibnamefont {Lemonnier}}, \bibinfo {author}
  {\bibfnamefont {Z.}~\bibnamefont {Li}}, \bibinfo {author} {\bibfnamefont
  {C.~A.}\ \bibnamefont {Muryn}}, \bibinfo {author} {\bibfnamefont {L.~I.}\
  \bibnamefont {Palmer}}, \bibinfo {author} {\bibfnamefont {Y.}~\bibnamefont
  {Song}}, \bibinfo {author} {\bibfnamefont {G.~F.}\ \bibnamefont {Whitehead}},
  \ and\ \bibinfo {author} {\bibfnamefont {R.~J.}\ \bibnamefont {Young}},\
  }\href@noop {} {\bibfield  {journal} {\bibinfo  {journal} {Nature}\ }\textbf
  {\bibinfo {volume} {588}},\ \bibinfo {pages} {429} (\bibinfo {year}
  {2020})}\BibitemShut {NoStop}%
\bibitem [{\citenamefont {Fosado}\ \emph {et~al.}(2024)\citenamefont {Fosado},
  \citenamefont {Michieletto},\ and\ \citenamefont {Martelli}}]{Fosado2024ice}%
  \BibitemOpen
  \bibfield  {author} {\bibinfo {author} {\bibfnamefont {Y.~A.~G.}\
  \bibnamefont {Fosado}}, \bibinfo {author} {\bibfnamefont {D.}~\bibnamefont
  {Michieletto}}, \ and\ \bibinfo {author} {\bibfnamefont {F.}~\bibnamefont
  {Martelli}},\ }\href {http://arxiv.org/abs/2406.09080} {\ ,\ \bibinfo {pages}
  {1} (\bibinfo {year} {2024})},\ \Eprint {http://arxiv.org/abs/2406.09080}
  {arXiv:2406.09080} \BibitemShut {NoStop}%
\bibitem [{\citenamefont {Lang}\ and\ \citenamefont {Kumar}(2021)}]{Lang2021}%
  \BibitemOpen
  \bibfield  {author} {\bibinfo {author} {\bibfnamefont {M.}~\bibnamefont
  {Lang}}\ and\ \bibinfo {author} {\bibfnamefont {K.~S.}\ \bibnamefont
  {Kumar}},\ }\href {\doibase 10.1021/acs.macromol.1c00718} {\bibfield
  {journal} {\bibinfo  {journal} {Macromolecules}\ }\textbf {\bibinfo {volume}
  {54}},\ \bibinfo {pages} {7021} (\bibinfo {year} {2021})}\BibitemShut
  {NoStop}%
\bibitem [{\citenamefont {Panoukidou}\ \emph {et~al.}(2024)\citenamefont
  {Panoukidou}, \citenamefont {Weir}, \citenamefont {Sorichetti}, \citenamefont
  {Fosado}, \citenamefont {Lenz},\ and\ \citenamefont
  {Michieletto}}]{Panoukidou2024}%
  \BibitemOpen
  \bibfield  {author} {\bibinfo {author} {\bibfnamefont {M.}~\bibnamefont
  {Panoukidou}}, \bibinfo {author} {\bibfnamefont {S.}~\bibnamefont {Weir}},
  \bibinfo {author} {\bibfnamefont {V.}~\bibnamefont {Sorichetti}}, \bibinfo
  {author} {\bibfnamefont {Y.~G.}\ \bibnamefont {Fosado}}, \bibinfo {author}
  {\bibfnamefont {M.}~\bibnamefont {Lenz}}, \ and\ \bibinfo {author}
  {\bibfnamefont {D.}~\bibnamefont {Michieletto}},\ }\href {\doibase
  10.1103/PhysRevResearch.6.023189} {\bibfield  {journal} {\bibinfo  {journal}
  {Physical Review Research}\ }\textbf {\bibinfo {volume} {6}},\ \bibinfo
  {pages} {1} (\bibinfo {year} {2024})}\BibitemShut {NoStop}%
\bibitem [{\citenamefont {Shimada}\ and\ \citenamefont
  {Yamakawa}(1984)}]{Shimada1984}%
  \BibitemOpen
  \bibfield  {author} {\bibinfo {author} {\bibfnamefont {J.}~\bibnamefont
  {Shimada}}\ and\ \bibinfo {author} {\bibfnamefont {H.}~\bibnamefont
  {Yamakawa}},\ }\href {\doibase 10.1021/ma00134a028} {\bibfield  {journal}
  {\bibinfo  {journal} {Macromolecules}\ }\textbf {\bibinfo {volume} {17}},\
  \bibinfo {pages} {689} (\bibinfo {year} {1984})}\BibitemShut {NoStop}%
\bibitem [{\citenamefont {Fischer}\ \emph {et~al.}(2015)\citenamefont
  {Fischer}, \citenamefont {Lang},\ and\ \citenamefont {Sommer}}]{Fischer2015}%
  \BibitemOpen
  \bibfield  {author} {\bibinfo {author} {\bibfnamefont {J.}~\bibnamefont
  {Fischer}}, \bibinfo {author} {\bibfnamefont {M.}~\bibnamefont {Lang}}, \
  and\ \bibinfo {author} {\bibfnamefont {J.~U.}\ \bibnamefont {Sommer}},\
  }\href {\doibase 10.1063/1.4933228} {\bibfield  {journal} {\bibinfo
  {journal} {Journal of Chemical Physics}\ }\textbf {\bibinfo {volume} {143}}
  (\bibinfo {year} {2015}),\ 10.1063/1.4933228}\BibitemShut {NoStop}%
\bibitem [{\citenamefont {Kreuzer}\ and\ \citenamefont
  {Cozzarelli}(1980)}]{Kreuzer1980}%
  \BibitemOpen
  \bibfield  {author} {\bibinfo {author} {\bibfnamefont {K.}~\bibnamefont
  {Kreuzer}}\ and\ \bibinfo {author} {\bibfnamefont {N.}~\bibnamefont
  {Cozzarelli}},\ }\href
  {http://www.sciencedirect.com/science/article/pii/0092867480902524}
  {\bibfield  {journal} {\bibinfo  {journal} {Cell}\ }\textbf {\bibinfo
  {volume} {20}},\ \bibinfo {pages} {245} (\bibinfo {year} {1980})}\BibitemShut
  {NoStop}%
\bibitem [{\citenamefont {Kim}\ \emph {et~al.}(2013)\citenamefont {Kim},
  \citenamefont {Kundukad}, \citenamefont {Allahverdi}, \citenamefont
  {Nordensk{\"{o}}ld}, \citenamefont {Doyle},\ and\ \citenamefont {{Van Der
  Maarel}}}]{Kim2013}%
  \BibitemOpen
  \bibfield  {author} {\bibinfo {author} {\bibfnamefont {Y.~S.}\ \bibnamefont
  {Kim}}, \bibinfo {author} {\bibfnamefont {B.}~\bibnamefont {Kundukad}},
  \bibinfo {author} {\bibfnamefont {A.}~\bibnamefont {Allahverdi}}, \bibinfo
  {author} {\bibfnamefont {L.}~\bibnamefont {Nordensk{\"{o}}ld}}, \bibinfo
  {author} {\bibfnamefont {P.~S.}\ \bibnamefont {Doyle}}, \ and\ \bibinfo
  {author} {\bibfnamefont {J.~R.~C.}\ \bibnamefont {{Van Der Maarel}}},\ }\href
  {\doibase 10.1039/c2sm27229f} {\bibfield  {journal} {\bibinfo  {journal}
  {Soft Matter}\ }\textbf {\bibinfo {volume} {9}},\ \bibinfo {pages} {1656}
  (\bibinfo {year} {2013})}\BibitemShut {NoStop}%
\bibitem [{\citenamefont {Krajina}\ \emph {et~al.}(2018)\citenamefont
  {Krajina}, \citenamefont {Zhu}, \citenamefont {Heilshorn},\ and\
  \citenamefont {Spakowitz}}]{Krajina2018}%
  \BibitemOpen
  \bibfield  {author} {\bibinfo {author} {\bibfnamefont {B.~A.}\ \bibnamefont
  {Krajina}}, \bibinfo {author} {\bibfnamefont {A.}~\bibnamefont {Zhu}},
  \bibinfo {author} {\bibfnamefont {S.~C.}\ \bibnamefont {Heilshorn}}, \ and\
  \bibinfo {author} {\bibfnamefont {A.~J.}\ \bibnamefont {Spakowitz}},\ }\href
  {\doibase 10.1103/PhysRevLett.121.148001} {\bibfield  {journal} {\bibinfo
  {journal} {Physical Review Letters}\ }\textbf {\bibinfo {volume} {121}},\
  \bibinfo {pages} {148001} (\bibinfo {year} {2018})}\BibitemShut {NoStop}%
\bibitem [{\citenamefont {Le}\ \emph {et~al.}(2023)\citenamefont {Le},
  \citenamefont {Wu}, \citenamefont {Lee}, \citenamefont {Bhatt}, \citenamefont
  {Inman}, \citenamefont {Berger},\ and\ \citenamefont {Wang}}]{Le2022}%
  \BibitemOpen
  \bibfield  {author} {\bibinfo {author} {\bibfnamefont {T.~T.}\ \bibnamefont
  {Le}}, \bibinfo {author} {\bibfnamefont {M.}~\bibnamefont {Wu}}, \bibinfo
  {author} {\bibfnamefont {J.~H.}\ \bibnamefont {Lee}}, \bibinfo {author}
  {\bibfnamefont {N.}~\bibnamefont {Bhatt}}, \bibinfo {author} {\bibfnamefont
  {J.~T.}\ \bibnamefont {Inman}}, \bibinfo {author} {\bibfnamefont {J.~M.}\
  \bibnamefont {Berger}}, \ and\ \bibinfo {author} {\bibfnamefont {M.~D.}\
  \bibnamefont {Wang}},\ }\href {\doibase 10.1038/s41589-022-01235-9}
  {\bibfield  {journal} {\bibinfo  {journal} {Nature Chemical Biology}\ }
  (\bibinfo {year} {2023}),\ 10.1038/s41589-022-01235-9}\BibitemShut {NoStop}%
\bibitem [{\citenamefont {Ubertini}\ and\ \citenamefont
  {Rosa}(2021)}]{Ubertini2021}%
  \BibitemOpen
  \bibfield  {author} {\bibinfo {author} {\bibfnamefont {M.~A.}\ \bibnamefont
  {Ubertini}}\ and\ \bibinfo {author} {\bibfnamefont {A.}~\bibnamefont
  {Rosa}},\ }\href {\doibase 10.1103/PhysRevE.104.054503} {\bibfield  {journal}
  {\bibinfo  {journal} {Physical Review E}\ }\textbf {\bibinfo {volume}
  {104}},\ \bibinfo {pages} {1} (\bibinfo {year} {2021})}\BibitemShut {NoStop}%
\bibitem [{\citenamefont {Vilgis}\ and\ \citenamefont
  {Otto}(1997)}]{Vilgis1997}%
  \BibitemOpen
  \bibfield  {author} {\bibinfo {author} {\bibfnamefont {T.~A.}\ \bibnamefont
  {Vilgis}}\ and\ \bibinfo {author} {\bibfnamefont {M.}~\bibnamefont {Otto}},\
  }\href {http://link.aps.org/doi/10.1103/PhysRevE.56.R1314} {\bibfield
  {journal} {\bibinfo  {journal} {Phys. Rev. E}\ }\textbf {\bibinfo {volume}
  {56}},\ \bibinfo {pages} {R1314} (\bibinfo {year} {1997})}\BibitemShut
  {NoStop}%
\bibitem [{\citenamefont {Lang}\ \emph {et~al.}(2014)\citenamefont {Lang},
  \citenamefont {Fischer}, \citenamefont {Werner},\ and\ \citenamefont
  {Sommer}}]{Lang2014a}%
  \BibitemOpen
  \bibfield  {author} {\bibinfo {author} {\bibfnamefont {M.}~\bibnamefont
  {Lang}}, \bibinfo {author} {\bibfnamefont {J.}~\bibnamefont {Fischer}},
  \bibinfo {author} {\bibfnamefont {M.}~\bibnamefont {Werner}}, \ and\ \bibinfo
  {author} {\bibfnamefont {J.~U.}\ \bibnamefont {Sommer}},\ }\href {\doibase
  10.1103/PhysRevLett.112.238001} {\bibfield  {journal} {\bibinfo  {journal}
  {Physical Review Letters}\ }\textbf {\bibinfo {volume} {112}},\ \bibinfo
  {pages} {1} (\bibinfo {year} {2014})}\BibitemShut {NoStop}%
\bibitem [{\citenamefont {Rubinstein}\ and\ \citenamefont
  {Colby}(2003)}]{Rubinsteinbook}%
  \BibitemOpen
  \bibfield  {author} {\bibinfo {author} {\bibfnamefont {M.}~\bibnamefont
  {Rubinstein}}\ and\ \bibinfo {author} {\bibfnamefont {H.~R.}\ \bibnamefont
  {Colby}},\ }\href@noop {} {\emph {\bibinfo {title} {{Polymer Physics}}}}\
  (\bibinfo  {publisher} {Oxford University Press},\ \bibinfo {year}
  {2003})\BibitemShut {NoStop}%
\bibitem [{Note2()}]{Note2}%
  \BibitemOpen
  \bibinfo {note} {There is no evidence that de Gennes and Sauvage knew of each
  other's work, despite being compatriots.}\BibitemShut {Stop}%
\bibitem [{\citenamefont {Dietrich-Buchecker}\ \emph
  {et~al.}(1983)\citenamefont {Dietrich-Buchecker}, \citenamefont {Sauvage},\
  and\ \citenamefont {Kintzinger}}]{Sauvage1983}%
  \BibitemOpen
  \bibfield  {author} {\bibinfo {author} {\bibfnamefont {C.~O.}\ \bibnamefont
  {Dietrich-Buchecker}}, \bibinfo {author} {\bibfnamefont {J.~P.}\ \bibnamefont
  {Sauvage}}, \ and\ \bibinfo {author} {\bibfnamefont {J.~P.}\ \bibnamefont
  {Kintzinger}},\ }\href {\doibase 10.1016/S0040-4039(00)94050-4} {\bibfield
  {journal} {\bibinfo  {journal} {Tetrahedron Letters}\ }\textbf {\bibinfo
  {volume} {24}},\ \bibinfo {pages} {5095} (\bibinfo {year}
  {1983})}\BibitemShut {NoStop}%
\bibitem [{\citenamefont {Hudson}\ and\ \citenamefont
  {Vinograd}(1967)}]{Hudson1967}%
  \BibitemOpen
  \bibfield  {author} {\bibinfo {author} {\bibfnamefont {B.}~\bibnamefont
  {Hudson}}\ and\ \bibinfo {author} {\bibfnamefont {J.}~\bibnamefont
  {Vinograd}},\ }\href@noop {} {\bibfield  {journal} {\bibinfo  {journal}
  {Nature}\ }\textbf {\bibinfo {volume} {216}},\ \bibinfo {pages} {647}
  (\bibinfo {year} {1967})}\BibitemShut {NoStop}%
\bibitem [{\citenamefont {Clayton}\ and\ \citenamefont
  {Vinograd}(1967)}]{Clayton1967}%
  \BibitemOpen
  \bibfield  {author} {\bibinfo {author} {\bibfnamefont {D.~A.}\ \bibnamefont
  {Clayton}}\ and\ \bibinfo {author} {\bibfnamefont {J.}~\bibnamefont
  {Vinograd}},\ }\href@noop {} {\bibfield  {journal} {\bibinfo  {journal}
  {Nature}\ }\textbf {\bibinfo {volume} {216}},\ \bibinfo {pages} {652}
  (\bibinfo {year} {1967})}\BibitemShut {NoStop}%
\bibitem [{\citenamefont {Fielden}\ \emph {et~al.}(2017)\citenamefont
  {Fielden}, \citenamefont {Leigh},\ and\ \citenamefont
  {Woltering}}]{Fielden2017}%
  \BibitemOpen
  \bibfield  {author} {\bibinfo {author} {\bibfnamefont {S.~D.}\ \bibnamefont
  {Fielden}}, \bibinfo {author} {\bibfnamefont {D.~A.}\ \bibnamefont {Leigh}},
  \ and\ \bibinfo {author} {\bibfnamefont {S.~L.}\ \bibnamefont {Woltering}},\
  }\href {\doibase 10.1002/anie.201702531} {\bibfield  {journal} {\bibinfo
  {journal} {Angewandte Chemie - International Edition}\ }\textbf {\bibinfo
  {volume} {56}},\ \bibinfo {pages} {11166} (\bibinfo {year}
  {2017})}\BibitemShut {NoStop}%
\bibitem [{\citenamefont {Schaufelberger}(2020)}]{Schaufelberger2020}%
  \BibitemOpen
  \bibfield  {author} {\bibinfo {author} {\bibfnamefont {F.}~\bibnamefont
  {Schaufelberger}},\ }\href {\doibase 10.1038/s42004-020-00433-7} {\bibfield
  {journal} {\bibinfo  {journal} {Communications Chemistry}\ }\textbf {\bibinfo
  {volume} {3}},\ \bibinfo {pages} {1} (\bibinfo {year} {2020})}\BibitemShut
  {NoStop}%
\bibitem [{\citenamefont {Zhao}\ \emph {et~al.}(2021)\citenamefont {Zhao},
  \citenamefont {Zhang}, \citenamefont {Zhao}, \citenamefont {Liu},
  \citenamefont {Liu}, \citenamefont {Li}, \citenamefont {Zhang}, \citenamefont
  {Bai}, \citenamefont {Yang},\ and\ \citenamefont {Yan}}]{Zhao2021}%
  \BibitemOpen
  \bibfield  {author} {\bibinfo {author} {\bibfnamefont {D.}~\bibnamefont
  {Zhao}}, \bibinfo {author} {\bibfnamefont {Z.}~\bibnamefont {Zhang}},
  \bibinfo {author} {\bibfnamefont {J.}~\bibnamefont {Zhao}}, \bibinfo {author}
  {\bibfnamefont {K.}~\bibnamefont {Liu}}, \bibinfo {author} {\bibfnamefont
  {Y.}~\bibnamefont {Liu}}, \bibinfo {author} {\bibfnamefont {G.}~\bibnamefont
  {Li}}, \bibinfo {author} {\bibfnamefont {X.}~\bibnamefont {Zhang}}, \bibinfo
  {author} {\bibfnamefont {R.}~\bibnamefont {Bai}}, \bibinfo {author}
  {\bibfnamefont {X.}~\bibnamefont {Yang}}, \ and\ \bibinfo {author}
  {\bibfnamefont {X.}~\bibnamefont {Yan}},\ }\href {\doibase
  10.1002/anie.202105620} {\bibfield  {journal} {\bibinfo  {journal}
  {Angewandte Chemie - International Edition}\ }\textbf {\bibinfo {volume}
  {60}},\ \bibinfo {pages} {16224} (\bibinfo {year} {2021})}\BibitemShut
  {NoStop}%
\bibitem [{\citenamefont {Kadic}\ \emph {et~al.}(2019)\citenamefont {Kadic},
  \citenamefont {Milton}, \citenamefont {van Hecke},\ and\ \citenamefont
  {Wegener}}]{Kadic2019}%
  \BibitemOpen
  \bibfield  {author} {\bibinfo {author} {\bibfnamefont {M.}~\bibnamefont
  {Kadic}}, \bibinfo {author} {\bibfnamefont {G.~W.}\ \bibnamefont {Milton}},
  \bibinfo {author} {\bibfnamefont {M.}~\bibnamefont {van Hecke}}, \ and\
  \bibinfo {author} {\bibfnamefont {M.}~\bibnamefont {Wegener}},\ }\href
  {\doibase 10.1038/s42254-018-0018-y} {\bibfield  {journal} {\bibinfo
  {journal} {Nature Reviews Physics}\ }\textbf {\bibinfo {volume} {1}},\
  \bibinfo {pages} {198} (\bibinfo {year} {2019})}\BibitemShut {NoStop}%
\bibitem [{\citenamefont {Diao}\ \emph {et~al.}(2012)\citenamefont {Diao},
  \citenamefont {Hinson}, \citenamefont {Kaplan}, \citenamefont {Vazquez},\
  and\ \citenamefont {Arsuaga}}]{Diao2012}%
  \BibitemOpen
  \bibfield  {author} {\bibinfo {author} {\bibfnamefont {Y.}~\bibnamefont
  {Diao}}, \bibinfo {author} {\bibfnamefont {K.}~\bibnamefont {Hinson}},
  \bibinfo {author} {\bibfnamefont {R.}~\bibnamefont {Kaplan}}, \bibinfo
  {author} {\bibfnamefont {M.}~\bibnamefont {Vazquez}}, \ and\ \bibinfo
  {author} {\bibfnamefont {J.}~\bibnamefont {Arsuaga}},\ }\href {\doibase
  10.1007/s00285-011-0438-0} {\bibfield  {journal} {\bibinfo  {journal} {J.
  Math. Biol.}\ }\textbf {\bibinfo {volume} {64}},\ \bibinfo {pages} {1087}
  (\bibinfo {year} {2012})}\BibitemShut {NoStop}%
\bibitem [{\citenamefont {Schmidt}\ and\ \citenamefont
  {Heckel}(2011)}]{Schmidt2011}%
  \BibitemOpen
  \bibfield  {author} {\bibinfo {author} {\bibfnamefont {T.~L.}\ \bibnamefont
  {Schmidt}}\ and\ \bibinfo {author} {\bibfnamefont {A.}~\bibnamefont
  {Heckel}},\ }\href {\doibase 10.1021/nl200303m} {\bibfield  {journal}
  {\bibinfo  {journal} {Nano Letters}\ }\textbf {\bibinfo {volume} {11}},\
  \bibinfo {pages} {1739} (\bibinfo {year} {2011})}\BibitemShut {NoStop}%
\bibitem [{\citenamefont {Rauscher}\ \emph {et~al.}(2020)\citenamefont
  {Rauscher}, \citenamefont {Schweizer}, \citenamefont {Rowan},\ and\
  \citenamefont {{De Pablo}}}]{Rauscher2020}%
  \BibitemOpen
  \bibfield  {author} {\bibinfo {author} {\bibfnamefont {P.~M.}\ \bibnamefont
  {Rauscher}}, \bibinfo {author} {\bibfnamefont {K.~S.}\ \bibnamefont
  {Schweizer}}, \bibinfo {author} {\bibfnamefont {S.~J.}\ \bibnamefont
  {Rowan}}, \ and\ \bibinfo {author} {\bibfnamefont {J.~J.}\ \bibnamefont {{De
  Pablo}}},\ }\href@noop {} {\bibfield  {journal} {\bibinfo  {journal}
  {Macromolecules}\ }\textbf {\bibinfo {volume} {53}},\ \bibinfo {pages} {3390}
  (\bibinfo {year} {2020})}\BibitemShut {NoStop}%
\bibitem [{\citenamefont {Luengo-M{\'{a}}rquez}\ \emph
  {et~al.}(2024)\citenamefont {Luengo-M{\'{a}}rquez}, \citenamefont {Assenza},\
  and\ \citenamefont {Micheletti}}]{Luengo-Marquez2024}%
  \BibitemOpen
  \bibfield  {author} {\bibinfo {author} {\bibfnamefont {J.}~\bibnamefont
  {Luengo-M{\'{a}}rquez}}, \bibinfo {author} {\bibfnamefont {S.}~\bibnamefont
  {Assenza}}, \ and\ \bibinfo {author} {\bibfnamefont {C.}~\bibnamefont
  {Micheletti}},\ }\href {http://arxiv.org/abs/2406.13561} {\ ,\ \bibinfo
  {pages} {1} (\bibinfo {year} {2024})},\ \Eprint
  {http://arxiv.org/abs/2406.13561} {arXiv:2406.13561} \BibitemShut {NoStop}%
\bibitem [{\citenamefont {Klotz}\ \emph {et~al.}(2024)\citenamefont {Klotz},
  \citenamefont {Anderson},\ and\ \citenamefont {Dimitriyev}}]{Klotz2024}%
  \BibitemOpen
  \bibfield  {author} {\bibinfo {author} {\bibfnamefont {A.~R.}\ \bibnamefont
  {Klotz}}, \bibinfo {author} {\bibfnamefont {C.~J.}\ \bibnamefont {Anderson}},
  \ and\ \bibinfo {author} {\bibfnamefont {M.~S.}\ \bibnamefont {Dimitriyev}},\
  }\href {http://arxiv.org/abs/2406.13590} {\bibfield  {journal} {\bibinfo
  {journal} {arxiv}\ } (\bibinfo {year} {2024})},\ \Eprint
  {http://arxiv.org/abs/2406.13590} {arXiv:2406.13590} \BibitemShut {NoStop}%
\bibitem [{\citenamefont {Polles}\ \emph {et~al.}(2016)\citenamefont {Polles},
  \citenamefont {Orlandini},\ and\ \citenamefont {Micheletti}}]{Polles2016}%
  \BibitemOpen
  \bibfield  {author} {\bibinfo {author} {\bibfnamefont {G.}~\bibnamefont
  {Polles}}, \bibinfo {author} {\bibfnamefont {E.}~\bibnamefont {Orlandini}}, \
  and\ \bibinfo {author} {\bibfnamefont {C.}~\bibnamefont {Micheletti}},\
  }\href@noop {} {\bibfield  {journal} {\bibinfo  {journal} {ACS Macro
  Letters}\ }\textbf {\bibinfo {volume} {5}},\ \bibinfo {pages} {931} (\bibinfo
  {year} {2016})}\BibitemShut {NoStop}%
\bibitem [{\citenamefont {Li}\ \emph {et~al.}(2021)\citenamefont {Li},
  \citenamefont {Chen},\ and\ \citenamefont {Choi}}]{Li2021}%
  \BibitemOpen
  \bibfield  {author} {\bibinfo {author} {\bibfnamefont {R.}~\bibnamefont
  {Li}}, \bibinfo {author} {\bibfnamefont {H.}~\bibnamefont {Chen}}, \ and\
  \bibinfo {author} {\bibfnamefont {J.~H.}\ \bibnamefont {Choi}},\ }\href
  {\doibase 10.1002/anie.202014729} {\bibfield  {journal} {\bibinfo  {journal}
  {Angewandte Chemie - International Edition}\ }\textbf {\bibinfo {volume}
  {60}},\ \bibinfo {pages} {7165} (\bibinfo {year} {2021})}\BibitemShut
  {NoStop}%
\bibitem [{\citenamefont {Yao}\ \emph {et~al.}(2020)\citenamefont {Yao},
  \citenamefont {Zhang}, \citenamefont {Wang}, \citenamefont {Peng},
  \citenamefont {Liu}, \citenamefont {Poppleton}, \citenamefont {{\v{S}}ulc},
  \citenamefont {Jiang}, \citenamefont {Liu}, \citenamefont {Gong},
  \citenamefont {Jing}, \citenamefont {Liu}, \citenamefont {Wang},
  \citenamefont {Liu}, \citenamefont {Fan},\ and\ \citenamefont
  {Yan}}]{Yao2020}%
  \BibitemOpen
  \bibfield  {author} {\bibinfo {author} {\bibfnamefont {G.}~\bibnamefont
  {Yao}}, \bibinfo {author} {\bibfnamefont {F.}~\bibnamefont {Zhang}}, \bibinfo
  {author} {\bibfnamefont {F.}~\bibnamefont {Wang}}, \bibinfo {author}
  {\bibfnamefont {T.}~\bibnamefont {Peng}}, \bibinfo {author} {\bibfnamefont
  {H.}~\bibnamefont {Liu}}, \bibinfo {author} {\bibfnamefont {E.}~\bibnamefont
  {Poppleton}}, \bibinfo {author} {\bibfnamefont {P.}~\bibnamefont
  {{\v{S}}ulc}}, \bibinfo {author} {\bibfnamefont {S.}~\bibnamefont {Jiang}},
  \bibinfo {author} {\bibfnamefont {L.}~\bibnamefont {Liu}}, \bibinfo {author}
  {\bibfnamefont {C.}~\bibnamefont {Gong}}, \bibinfo {author} {\bibfnamefont
  {X.}~\bibnamefont {Jing}}, \bibinfo {author} {\bibfnamefont {X.}~\bibnamefont
  {Liu}}, \bibinfo {author} {\bibfnamefont {L.}~\bibnamefont {Wang}}, \bibinfo
  {author} {\bibfnamefont {Y.}~\bibnamefont {Liu}}, \bibinfo {author}
  {\bibfnamefont {C.}~\bibnamefont {Fan}}, \ and\ \bibinfo {author}
  {\bibfnamefont {H.}~\bibnamefont {Yan}},\ }\href {\doibase
  10.1038/s41557-020-0539-8} {\bibfield  {journal} {\bibinfo  {journal} {Nature
  Chemistry}\ }\textbf {\bibinfo {volume} {12}},\ \bibinfo {pages} {1067}
  (\bibinfo {year} {2020})}\BibitemShut {NoStop}%
\bibitem [{\citenamefont {List}\ \emph {et~al.}(2016)\citenamefont {List},
  \citenamefont {Falgenhauer}, \citenamefont {Kopperger}, \citenamefont
  {Pardatscher},\ and\ \citenamefont {Simmel}}]{List2016}%
  \BibitemOpen
  \bibfield  {author} {\bibinfo {author} {\bibfnamefont {J.}~\bibnamefont
  {List}}, \bibinfo {author} {\bibfnamefont {E.}~\bibnamefont {Falgenhauer}},
  \bibinfo {author} {\bibfnamefont {E.}~\bibnamefont {Kopperger}}, \bibinfo
  {author} {\bibfnamefont {G.}~\bibnamefont {Pardatscher}}, \ and\ \bibinfo
  {author} {\bibfnamefont {F.~C.}\ \bibnamefont {Simmel}},\ }\href {\doibase
  10.1038/ncomms12414} {\bibfield  {journal} {\bibinfo  {journal} {Nature
  Communications}\ }\textbf {\bibinfo {volume} {7}},\ \bibinfo {pages} {1}
  (\bibinfo {year} {2016})}\BibitemShut {NoStop}%
\bibitem [{\citenamefont {Valero}\ \emph {et~al.}(2019)\citenamefont {Valero},
  \citenamefont {Centola}, \citenamefont {Ma}, \citenamefont {{\v{S}}kugor},
  \citenamefont {Yu}, \citenamefont {Haydell}, \citenamefont {Keppner},\ and\
  \citenamefont {Famulok}}]{Valero2019}%
  \BibitemOpen
  \bibfield  {author} {\bibinfo {author} {\bibfnamefont {J.}~\bibnamefont
  {Valero}}, \bibinfo {author} {\bibfnamefont {M.}~\bibnamefont {Centola}},
  \bibinfo {author} {\bibfnamefont {Y.}~\bibnamefont {Ma}}, \bibinfo {author}
  {\bibfnamefont {M.}~\bibnamefont {{\v{S}}kugor}}, \bibinfo {author}
  {\bibfnamefont {Z.}~\bibnamefont {Yu}}, \bibinfo {author} {\bibfnamefont
  {M.~W.}\ \bibnamefont {Haydell}}, \bibinfo {author} {\bibfnamefont
  {D.}~\bibnamefont {Keppner}}, \ and\ \bibinfo {author} {\bibfnamefont
  {M.}~\bibnamefont {Famulok}},\ }\href {\doibase 10.1038/s41596-019-0198-7}
  {\emph {\bibinfo {title} {Nature Protocols}}},\ Vol.~\bibinfo {volume} {14}\
  (\bibinfo {year} {2019})\ pp.\ \bibinfo {pages} {2818--2855}\BibitemShut
  {NoStop}%
\bibitem [{\citenamefont {Peil}\ \emph {et~al.}(2020)\citenamefont {Peil},
  \citenamefont {Zhan},\ and\ \citenamefont {Liu}}]{Peil2020}%
  \BibitemOpen
  \bibfield  {author} {\bibinfo {author} {\bibfnamefont {A.}~\bibnamefont
  {Peil}}, \bibinfo {author} {\bibfnamefont {P.}~\bibnamefont {Zhan}}, \ and\
  \bibinfo {author} {\bibfnamefont {N.}~\bibnamefont {Liu}},\ }\href {\doibase
  10.1002/smll.201905987} {\bibfield  {journal} {\bibinfo  {journal} {Small}\
  }\textbf {\bibinfo {volume} {16}},\ \bibinfo {pages} {1} (\bibinfo {year}
  {2020})}\BibitemShut {NoStop}%
\bibitem [{\citenamefont {Biffi}\ \emph {et~al.}(2013)\citenamefont {Biffi},
  \citenamefont {Cerbino}, \citenamefont {Bomboi}, \citenamefont {Paraboschi},
  \citenamefont {Asselta}, \citenamefont {Sciortino},\ and\ \citenamefont
  {Bellini}}]{Biffi2013}%
  \BibitemOpen
  \bibfield  {author} {\bibinfo {author} {\bibfnamefont {S.}~\bibnamefont
  {Biffi}}, \bibinfo {author} {\bibfnamefont {R.}~\bibnamefont {Cerbino}},
  \bibinfo {author} {\bibfnamefont {F.}~\bibnamefont {Bomboi}}, \bibinfo
  {author} {\bibfnamefont {E.~M.}\ \bibnamefont {Paraboschi}}, \bibinfo
  {author} {\bibfnamefont {R.}~\bibnamefont {Asselta}}, \bibinfo {author}
  {\bibfnamefont {F.}~\bibnamefont {Sciortino}}, \ and\ \bibinfo {author}
  {\bibfnamefont {T.}~\bibnamefont {Bellini}},\ }\href@noop {} {\bibfield
  {journal} {\bibinfo  {journal} {Proceedings of the National Academy of
  Sciences of the United States of America}\ }\textbf {\bibinfo {volume}
  {110}},\ \bibinfo {pages} {15633} (\bibinfo {year} {2013})}\BibitemShut
  {NoStop}%
\bibitem [{\citenamefont {Conrad}\ \emph {et~al.}(2019)\citenamefont {Conrad},
  \citenamefont {Kennedy}, \citenamefont {Fygenson},\ and\ \citenamefont
  {Saleh}}]{Conrad2019a}%
  \BibitemOpen
  \bibfield  {author} {\bibinfo {author} {\bibfnamefont {N.}~\bibnamefont
  {Conrad}}, \bibinfo {author} {\bibfnamefont {T.}~\bibnamefont {Kennedy}},
  \bibinfo {author} {\bibfnamefont {D.~K.}\ \bibnamefont {Fygenson}}, \ and\
  \bibinfo {author} {\bibfnamefont {O.~A.}\ \bibnamefont {Saleh}},\ }\href@noop
  {} {\bibfield  {journal} {\bibinfo  {journal} {Proceedings of the National
  Academy of Sciences of the United States of America}\ }\textbf {\bibinfo
  {volume} {116}},\ \bibinfo {pages} {7238} (\bibinfo {year}
  {2019})}\BibitemShut {NoStop}%
\bibitem [{\citenamefont {Neophytou}\ \emph {et~al.}(2022)\citenamefont
  {Neophytou}, \citenamefont {Chakrabarti},\ and\ \citenamefont
  {Sciortino}}]{Neophytou2022}%
  \BibitemOpen
  \bibfield  {author} {\bibinfo {author} {\bibfnamefont {A.}~\bibnamefont
  {Neophytou}}, \bibinfo {author} {\bibfnamefont {D.}~\bibnamefont
  {Chakrabarti}}, \ and\ \bibinfo {author} {\bibfnamefont {F.}~\bibnamefont
  {Sciortino}},\ }\href@noop {} {\bibfield  {journal} {\bibinfo  {journal}
  {Nature Physics}\ }\textbf {\bibinfo {volume} {18}},\ \bibinfo {pages} {1248}
  (\bibinfo {year} {2022})}\BibitemShut {NoStop}%
\bibitem [{\citenamefont {Nitiss}(1998)}]{Nitiss1998}%
  \BibitemOpen
  \bibfield  {author} {\bibinfo {author} {\bibfnamefont {J.~L.}\ \bibnamefont
  {Nitiss}},\ }\href {\doibase 10.1002/0471141755.ph0303s00} {\bibfield
  {journal} {\bibinfo  {journal} {Current Protocols in Pharmacology}\ }\textbf
  {\bibinfo {volume} {00}},\ \bibinfo {pages} {373} (\bibinfo {year}
  {1998})}\BibitemShut {NoStop}%
\bibitem [{\citenamefont {Waraich}\ \emph {et~al.}(2020)\citenamefont
  {Waraich}, \citenamefont {Jain}, \citenamefont {Colloms}, \citenamefont
  {Stark}, \citenamefont {Burton},\ and\ \citenamefont
  {Maxwell}}]{Waraich2020}%
  \BibitemOpen
  \bibfield  {author} {\bibinfo {author} {\bibfnamefont {N.~F.}\ \bibnamefont
  {Waraich}}, \bibinfo {author} {\bibfnamefont {S.}~\bibnamefont {Jain}},
  \bibinfo {author} {\bibfnamefont {S.~D.}\ \bibnamefont {Colloms}}, \bibinfo
  {author} {\bibfnamefont {W.~M.}\ \bibnamefont {Stark}}, \bibinfo {author}
  {\bibfnamefont {N.~P.}\ \bibnamefont {Burton}}, \ and\ \bibinfo {author}
  {\bibfnamefont {A.}~\bibnamefont {Maxwell}},\ }\href {\doibase
  10.2144/btn-2020-0059} {\bibfield  {journal} {\bibinfo  {journal}
  {BioTechniques}\ }\textbf {\bibinfo {volume} {69}},\ \bibinfo {pages} {357}
  (\bibinfo {year} {2020})}\BibitemShut {NoStop}%
\end{thebibliography}%
\end{document}